\documentclass[12pt,a4paper,english]{article}

\usepackage[bindingoffset=0.3cm,textheight=22.5cm,hdivide={2.7cm,*,2.7cm}, vdivide={*,22cm,*}]{geometry}
\usepackage[bookmarksnumbered=true,breaklinks=true]{hyperref}

\usepackage{amsmath,amsfonts,amssymb,babel,slashed,graphicx,color}

\usepackage[numbers,square,comma,sort&compress]{natbib}

\IfFileExists{dsfont.sty}
	{\usepackage{dsfont}
         \newcommand{\id}{\mathds{1}}}
	{\typeout{Package dsfont.sty was not found, using alternative macros.}
         \let\mathds=\mathbb
         \newcommand{\id}{\mbox{1 \kern-.59em {\rm l}}}}
\let\one=\id

\makeatletter

         \@addtoreset{equation}{section}
\makeatother

\newcommand{\nocontentsline}[3]{}
\newcommand{\tocless}[3]{\bgroup\let\addcontentsline=\nocontentsline#1{#2}#3\egroup}

\newcommand{\qed}{\nobreak \ifvmode \relax \else
      \ifdim\lastskip<1.5em \hskip-\lastskip
      \hskip1.5em plus0em minus0.5em \fi \nobreak
      \vrule height0.75em width0.5em depth0.25em\fi}

\newcommand{\be}{\begin{equation}}
\newcommand{\ee}{\end{equation}}
\newcommand{\eq}[1]{(\ref{#1})}

\def\nn{\nonumber}
\def\bea{\begin{eqnarray}}
\def\eea{\end{eqnarray}}
\def\obar{\overline}

\def\beqa{\begin{eqnarray}} 
\def\eeqa{\end{eqnarray}} 
\def\beq{\begin{equation}} 
\def\eeq{\end{equation}} 

\def\a{\alpha}          
\def\b{\beta}           
\def\c{\gamma}  
\def\d{\delta}

\def\g{\gamma} 

\def\k{\kappa}
\def\l{\lambda} \def\L{\Lambda}

\def\s{\sigma}  
\def\t{\tau}

  \def\cC{{\cal C}}
\def\cD{{\cal D}}  
 \def\cH{{\cal H}} \def\cI{{\cal I}}
\def\cJ{{\cal J}} \def\cK{{\cal K}} 
 \def\cN{{\cal N}} \def\cO{{\cal O}}
  
 \def\cT{{\cal T}} 
 \def\cW{{\cal W}}

\def\mg{\mathfrak{g}}

\newcommand{\R}{\mathds{R}}
\newcommand{\C}{\mathds{C}}
\newcommand{\N}{\mathds{N}}
\newcommand{\Z}{\mathds{Z}}

\newcommand{\msu}{\mathfrak{s}\mathfrak{u}}

\newcommand{\mmu}{\mathfrak{u}}

\def\bit{\begin{itemize}}
\def\eit{\end{itemize}}

\def\({\left(}
\def\){\right)}
\def\diag{\mbox{diag}}

\def\d{\delta}

 \def\del{\partial}

\newcommand{\tr}{\mbox{tr}}

\def\bcomment#1{}

\newcommand{\Di}{{\slashed{D}}}

\newcommand{\co}[2]{[#1,#2]}						
\renewcommand{\a}{\alpha}
\renewcommand{\b}{\beta}
\renewcommand{\d}{\delta}

\renewcommand{\l}{\lambda}
\renewcommand{\t}{\tau}

\newcommand{\w}{\omega}

\renewcommand{\L}{\Lambda}

\DeclareMathOperator{\Tr}{Tr}

 \allowdisplaybreaks[3]

\begin{document}

\renewcommand{\title}[1]{\vspace{10mm}\noindent{\Large{\bf
#1}}\vspace{8mm}} \newcommand{\authors}[1]{\noindent{\large
#1}\vspace{5mm}} \newcommand{\address}[1]{{\itshape #1\vspace{2mm}}}


\begin{flushright}
UWThPh-2014-25 
\end{flushright}

\begin{center}

\title{ \Large  Spinning squashed extra dimensions and chiral gauge theory 
from ${\cal N}=4$ SYM}

\vskip 3mm

\authors{Harold C. Steinacker{\footnote{harold.steinacker@univie.ac.at}}
}
 
\vskip 3mm

 \address{ 

{\it Faculty of Physics, University of Vienna\\
Boltzmanngasse 5, A-1090 Vienna, Austria  }  
  }

\vskip 1.4cm

\textbf{Abstract}

\end{center}

 New  solutions of $SU(N)$ ${\cal N}=4$ SYM on $\mathbb{R}^4$ interpreted as 
 spinning self-intersecting extra dimensions are discussed. 
 Remarkably, these backgrounds lead to a low-energy sector with 3 generations of 
 chiral fermions coupled to scalar and gauge fields, with standard Lorentz-invariant kinematics.
 This sector arises from zero modes localized on the rotation axes, 
 which are oblivious to the background rotation. 
 The remaining modes are not described by a Lorentz-invariant  
 field theory and are mostly ``heavy'', but there is one sextet of tachyonic excitations.
 Assuming that the latter get stabilized e.g. by  quantum effects, 
  we argue that different rotation frequencies would induce a VEV for 
 some of the low-energy scalar fields.
We discuss configurations which may lead to a low-energy physics not far from  the 
broken phase of the standard model.

\vskip 1.4cm

\textbf{keywords:}  fuzzy extra dimensions; $N=4$ super-Yang-Mills;
fuzzy spaces;
rotating branes

\newpage

\tableofcontents

\section{Introduction}\label{sec:background}

Simplicity has always been a central guiding principle in theoretical physics.
In the theory of fundamental interactions, 
this leads naturally to the idea of grand unified models. Another idea is that 
a simple higher-dimensional theory, such as string theory, might explain the rich
low-energy phenomenology in terms 
of some compactification $\R^4 \times \cK$. 
Remarkably, both ideas can be combined and realized
within 4-dimensional gauge theory, through a geometrical version of the Higgs mechanism. 
For example, starting with $\cN=4$ super-Yang-Mills (SYM) amended by a suitable cubic potential,
the behavior of a higher-dimensional theory on $\R^4 \times \cK_N$ may emerge
in a non-trivial vacuum, within a certain range of energies. 
Here $\cK_N$ is some approximation to a compact space $\cK \subset\R^6$, 
such as a fuzzy sphere, or some more complicated fuzzy manifold. 
This mechanism of dynamically generating fuzzy extra dimensions has been studied in various guises and 
examples, see e.g. \cite{DuboisViolette:1989at,Myers:1999ps,Polchinski:2000uf,Berenstein:2000ux,
ArkaniHamed:2001nc,Andrews:2005cv,Aschieri:2006uw,Steinacker:2007ay,
Chatzistavrakidis:2009ix,Chatzistavrakidis:2010xi,Kurkcuoglu:2010sn,Furuuchi:2011px}. 
It allows to obtain the attractive features and the structure of a higher-dimensional
theory starting from a simple 4-dimensional gauge theory.

Although  in basic examples the resulting low-energy physics is somewhat academic, 
progress has been made recently towards 
more interesting low-energy behavior. In particular, chiral fermions can be obtained
for backgrounds which locally span all 6 internal dimensions in $\cN=4$ SYM, 
e.g. via intersecting branes \cite{Chatzistavrakidis:2011gs,Steinacker:2014fja}, or other related mechanisms 
 \cite{Chatzistavrakidis:2010xi,Nishimura:2013moa,Aoki:2014cya}.
A particularly interesting and non-trivial example was found in \cite{Steinacker:2014lma}, 
where $\cK_N \cong \cC_N[\mu]$ is a fuzzy version of a squashed (or projected) self-intersecting
coadjoint orbit of $SU(3)$. This leads to
3 generations of chiral fermions in the low-energy theory, localized at
the origin of the extra dimensions.
These solutions can be stabilized by a  cubic soft SUSY breaking term in the potential,
which is added  by hand. This may seem unavoidable, but it spoils
the simplicity and the special role of $\cN=4$ SYM. 
Analogous solutions arise also in the IKKT or IIB matrix model \cite{Ishibashi:1996xs},
where the  cubic terms are more problematic since logarithmic UV divergences
typically lead to UV/IR mixing.

In the present paper, we show that the $\R^4 \times \cC_N[\mu]$ solutions arise even 
in pure vanilla $\cN=4$ SYM without any SUSY-breaking potential, if the compact space is allowed to spin.
This corresponds to a condensate of $SO(6)$ currents with maximal rank.
While there are  many spinning background solutions in $\cN=4$ SYM and related
matrix models (see e.g. \cite{Steinacker:2011wb,Polychronakos:2013fma}), 
they typically lead to significant 
breaking of Lorentz invariance  with unacceptable low-energy
kinematics, in particular for interesting cases with chiral fermions 
at intersecting branes \cite{meandJochen-unpub}. 
The miraculous property of the squashed branes $\cC_N[\mu]$  is that they
lead to a low-energy sector 
consisting of three generations of bosonic and fermionic chiral zero modes, which are oblivious to 
the background rotation, and governed by a standard Poincare-invariant 
field theory at low energies. 
These bosonic and fermionic zero modes naturally form chiral supermultiplets.
This ``miracle'' can be understood  by the fact that the chiral fermionic 
and bosonic zero modes arise at the intersection of the different sheets of the fuzzy 
brane at the origin  \cite{Steinacker:2014lma}, which is the center of rotation.
Besides these zero modes, there is a sector of deformed fluctuation modes which couple to the 
background rotation, and which are not described by an ordinary field theory.
We provide a formalism which allows to describe these fluctuations explicitly.
A numerical analysis indicates that most of these modes
are stable and ``heavy'', however there is a sextet of (slightly) unstable modes  i.e. resonances. 
In this paper, we simply assume that this sector can be stabilized somehow (e.g. by quantum 
corrections or by modifications of the background), and 
explore the possible consequences. 
Clearly this instability must be addressed in more detail elsewhere.

Studying the zero modes around these backgrounds leads to further intriguing observations.
If the three rotation frequencies $\omega_i$ are slightly different, then some of these 
bosonic zero modes acquire a negative mass, inducing further 
symmetry breaking; they will hence be called ``would-be zero modes''.
The scalar field equations  indeed admit solutions 
where some of these modes are switched on, with scale set by the {\em difference} of the $\omega_i$. 
This suggests an interesting mechanism for introducing a hierarchy of scales into the model. 
However, these solutions are only approximate neglecting the  
gauge fields, and more complete solutions remain to be found.

To get some insight into the low-energy physics which may emerge on such a background,
we start exploring the (rather complicated) structure of the Yukawa couplings 
between the low-energy Higgs modes and the chiral fermions, as well as the couplings between the 
low-energy fermionic currents and the  gauge fields.
This is particularly interesting for stacks of such branes.
The resulting pattern of fermions linking a stack of point branes 
attached to $\cC_N[\mu]$ branes leads quite naturally to the  matter content of the 
standard model coupled to the appropriate gauge fields, with all the correct quantum numbers.
It turns out that only a special subset of the Kaluza-Klein tower of gauge fields
couples to the low-energy currents, which includes in particular a pair 
of ``chiral'' gauge fields with distinct coupling to different chiral fermion modes
attached to the brane, independent of generation. 
Additional fields such as various gauginos and higher Kaluza-Klein modes also arise, 
and no claim on the physical viability is made. Nevertheless, it is 
 striking that one can arrive quite naturally in the vicinity of the 
 standard model (in the broken phase), reproducing 
precisely all its odd quantum numbers and even ``predicting'' the number of generations.

Although the underlying assumption of a spinning background may seem strange at first, it is in a sense 
quite generic. In fact, the assumption of having precisely zero background currents 
is non-generic, hence the present discussion is quite natural.
The  solutions under consideration simply amount to a
$SO(6)$ $R$-current condensate with full rank.
However,  the fate of the resonance modes  must be 
addressed in more detailed and resolved, in order to establish a reliable basis for further work.
In spite of this issue, the accumulation of little miracles in the most symmetric of all field theories
$\cN=4$ SYM is certainly remarkable, and 
lends to the hope that these solutions may point the way towards actual physics.

\section{Rotating branes in  $\cN=4$ SYM}

The action of $\cN=4$ $SU(N)$ SYM is organized most transparently 
in terms of 10-dimensional SYM reduced to 4 dimensions:
\begin{align}
 S_{\rm YM}
&=  \int d^4 x \ \frac 1{4g_{}^2} \tr\Big(-F^{\mu\nu} F_{\mu\nu} 
 - 2  D^\mu \Phi^a D_\mu \Phi_a +  [\Phi^a,\Phi^b][\Phi_a,\Phi_b] \Big)\nn\\
&\qquad \qquad +  \tr\Big(\bar\Psi\g^{\mu} i D_\mu \Psi + \bar\Psi\Gamma^a [\Phi_a,\Psi]\Big) .
\label{N=4SYM}
\end{align}
Here $F_{\mu\nu}$ is the field strength, $D_\mu = \del_\mu - i [A_\mu,.]$ the covariant derivative,
$\Phi^a,\ a \in \{1,2,4,5,6,7\}$ are 6 scalar 
fields\footnote{The unusual numbering of the indices anticipates their relation with $\msu(3)$ generators
in the following.}, 
$\Psi$ is a matrix-valued Majorana-Weyl spinor of $SO(9,1)$ dimensionally  
reduced to 4-dimensions, and $\Gamma^a$ arise from the 10-dimensional gamma matrices.
 All fields transform in the adjoint of the $SU(N)$ gauge symmetry, and
 the coupling constant $g$ is absorbed in the scalar fields $\Phi$.
It will be useful to work with dimensionless scalar fields labeled by the roots $\pm\a_i$ of $\msu(3)$, 
\begin{align}
 \Phi_\a =   m X_\a, \qquad \a\in \cI =  \{\pm\a_i,\ i=1,2,3 \}
\end{align}
where  $m$ has the dimension of a mass.
These $\a\in \cI$
 are viewed as points in $\R^2$ forming a hexagon, cf. figure \ref{fig:spin-states-roots}.
Explicitly,
\begin{align}
 X_1^\pm &= \frac 12(X_4\pm i X_5) \equiv X_{\pm \a_1}, \nn\\
 X_2^\pm &= \frac 12(X_6\mp i X_7)  \equiv X_{\pm \a_2}, \nn\\ 
 X_3^\pm &= \frac 12(X_1\pm i X_2)  \equiv X_{\pm \a_3} .
 \label{X-T-definition}
\end{align}
The potential for the scalar  fields can then be written as
\begin{align}
 V[X] &= - \frac {m^4}{4 g^2} \tr \Big(\sum_{\a,\b\in\cI}\co{X_\a}{X_\b}\co{X^\a}{X^\b}\, \Big)  
\end{align}
where
\begin{align}
 X^\a = 2 X_{-\a} .
\end{align}
Including variations around some background
\begin{align}
 X_\a \to X_\a + \phi_\a
\end{align}
the potential can be written as
\begin{align}
 V(X+\phi) &= \frac{m^4}{g^2}\tr \Big(-\frac 14 [X_\a,X_\b][X^\a,X^\b]
  + \phi_\a\Box_X X^\a  + X_\a\Box_\phi \phi^\a  \nn\\
  &\qquad + \frac 12\phi^\a\Big(\Box_X\d_\a^\b + 2 [[X_\a,X^\b],.] \Big)\phi_\b  - \frac 12 f^2
    - \frac 14 [\phi_\a,\phi_\b][\phi^\a,\phi^\b] \Big) .
 \label{full-potential}
\end{align}
Here
\begin{align}
 \Box_X  = \sum_{a\in\cI} [X_\a,[X^\a,.]]
 = 2([X_j^+, [X_j^-,.]] + [X_j^-, [X_j^+,.]])
 \label{matrix-Box}
\end{align}
and similarly $\Box_\phi$, and 
\begin{align}
 f = i[\phi_\a,X^\a]  \ 
 \label{gauge-fixing}
\end{align}
can be viewed as gauge-fixing function in extra dimensions.
Therefore the equations of motion (eom) are
\begin{align}
 (\Box_4 + m^2 \Box_X) X_\a  = 0
 \label{eom-roots}
\end{align}
where $\Box_4 = -D_\mu D^\mu$.
The quadratic terms $V_2(\phi)$ governing the fluctuations in $\phi^a$ is
\begin{align}
 V_2(\phi) & = \frac{m^4}{2g^2} \tr\Big(\phi_\a  \big(\Box_X\d^\a_\b +  2 [[X^\a, X_\b],. \, ] 
  - [X^\a,[X_\b,.]]\big) \phi^\b \Big) \ .
 \label{fluctuation-potential-complex}
\end{align}
We will often drop the last term, assuming the gauge $f=0$.

\subsection{Rotating squashed brane solutions}
\label{sec:rotating-solutions}

We recall the construction of the squashed fuzzy
coadjoint orbits $\cC_N[\mu]$ with singular embedding in $\R^6$ \cite{Steinacker:2014lma}. 
Let  $T_a,\ a=1,...,8$ be generators of the Lie algebra $\msu(3)$, 
with structure constants
\begin{align}
 [T_a,T_b] = ic_{abc} T_c
 \label{structure-const}
\end{align}
 canonically normalized with the Killing form
\begin{align}
\k_{ab} = (T_a,T_b)  =  2 \d_{ab} .
\label{killing}
\end{align}
The Cartan subalgebra is spanned by the  
two  Cartan generators $H_3 \equiv T_3$ and $H_8 \equiv T_8$.
The remaining generators $T_\a,\ \a\in\cI$
are combined into the ladder or root generators 
\begin{align}
 T_1^\pm &= \frac 12(T_4\pm i T_5) \equiv T_{\pm \a_1}, \nn\\
 T_2^\pm &= \frac 12(T_6\mp i T_7)\equiv T_{\pm \a_2},  \nn\\ 
 T_3^\pm &= \frac 12(T_1\pm i T_2) = \pm [T_1^\pm,T_2^\pm] \equiv T_{\pm \a_3} , 
 \label{X-T-definition}
\end{align}
where $\a_{1}, \a_2$ are the simple roots and $\a_3 = \a_1+\a_2$,
see figure \ref{fig:spin-states-roots}.
Then the Lie algebra relations can be written in the Cartan--Weyl form  as
\begin{align}
 [T_\a,T_\b] &= \pm T_{\a+\b}, \qquad 0 \neq \a+\b\in \cI  \nn\\
 [T_{\a_i},T_{-\a_i}] &= H_{i} \nn\\
 [H,T_\a] &= \a(H) T_\a \ 
 \label{Cartan-Weyl}
\end{align}
or explicitly
 \begin{align}
 [T_i^+,T_i^-] &=  H_{i}, \qquad i = 1,2,3  \nn\\
 [T_1^+,T_2^+] &=  T_3^+  \nn\\
 [T_1^+,T_3^-] &= - T_2^-  \nn\\
 [T_2^+,T_3^-] &=  T_1^-  \nn\\
 [T_1^+,T_2^-] &= [T_2^+,T_3^+] = [T_1^+,T_3^+]  = 0 .
\end{align}
We recall that the Cartan generators are canonically associated to weights $\a\in\mg_0^*$,
such that $H_\a |M\rangle  = (\a,M)|M\rangle$ for  weight states $|M\rangle$ in any representation;
in particular $H_i = H_{\a_i}$.
We also recall the Weyl group $\cW$, which is generated by the reflections in weight space 
along the  roots $\a_i$.

Let $\cH_{\mu}$ be the  irreducible representation\footnote{The fundamental representation 
with $\mu=\L_1$ corresponds to the  Gell-Mann matrices 
$\l_a = \pi_{(1,0)}(T_a)$.} 
with  highest weight $\mu = n_1 \L_1 + n_2 \L_2$, where 
$\L_i$ are the fundamental weights of $\msu(3)$.
For any $n_1,n_2\in\N$, this provides us with 6 hermitian matrices 
$X^a_{(\mu)} = \pi_\mu(T^a)$, or equivalently
\begin{align}
 X_{\pm\a_i} \equiv X_i^\pm = \pi_\mu(T_i^\pm), \qquad i=1,2,3 
\end{align}
(we will drop the subscript $\mu$ from now on).
Note that the Cartan generators are {\em not} included in the $X^a$.
As explained in \cite{Steinacker:2014lma}, the $X^a$
can be viewed as non-commutative embedding functions
$X^a \sim x^a: \cC[\mu] \hookrightarrow \R^8 \to  \R^6$, where $\cC[\mu]$ is a coadjoint orbit of $SU(3)$.
This defines squashed fuzzy $\cC_N[\mu]$, 
interpreted as noncommutative brane with squashed embedding in $\R^6$.
For $n_1=n_2$, these are projections of fuzzy $\C P^2_N$ \cite{Grosse:1999ci}.

As they stand, these matrices $X^a$ are of course not solutions of $\cN=4$ SYM. 
One possibility is  to add a cubic soft SUSY breaking term to the potential,
such that they are solutions \cite{Steinacker:2014lma}.
Here we pursue a different possibility, following the observation \cite{Steinacker:2011wb} that 
such branes can often be stabilized by rotation. 
Thus consider the following ansatz corresponding to spinning
squashed fuzzy $\cC_N[\mu]$ branes
\begin{align}\fbox{$ \ 
 X_i^\pm = r_i e^{\pm i \omega_{i} x}\, \pi_\mu(T_i^\pm)
  \ $}
  \label{rotating-brane}
\end{align}
for $i=1,2,3$, and $\omega x \equiv  \omega_{i,\mu} x^\mu$.
This amounts to a $x$-dependent $SO(6)$ rotation, or a rotating plane wave.
We allow for different (dimensionless) radii $r_i$ and rotation frequency vectors $\omega_{i,\mu}$.
Using the  Lie algebra relations, we compute
\begin{align}
 \frac 12 \Box_X X_i^+ &= \sum_j ([X_j^+,[X_j^-,X_i^+]] + [X_j^-,[X_j^+,X_i^+]]) \nn\\
 &=  r_i^2 ([T_i^+,[T_i^-,X_i^+]] + \sum_{j\neq i} r_j^2 [T_j^-,[T_j^+,X_i^+]]) \nn\\
 &= \big(2 r_i^2 + \sum_{j\neq i}  r_j^2\big) X_i^+ .
\end{align}
 Similarly,
 \begin{align}
  \Box_X X_1^\pm &= 2(2 r_1^2 +r_2^2 + r_3^2) X_1^\pm, \nn\\
  \Box_X X_2^\pm &= 2(r_1^2 + 2 r_2^2 + r_3^2) X_2^\pm, \nn\\
  \Box_X X_3^\pm &= 2(r_1^2 + r_2^2 + 2r_3^2) X_3^\pm .
  \label{Box-X-explicit}
 \end{align}
Hence the  eom \eq{eom-roots} become 
\begin{align}
 \omega_1^2 = -2m^2 (2 r_1^2 +r_2^2 + r_3^2)   \nn\\
 \omega_2^2 = -2m^2 (r_1^2 +2 r_2^2 + r_3^2)  \nn\\
 \omega_3^2 = -2m^2 (r_1^2 + r_2^2 + 2r_3^2)
 \label{eom-roots-explicit}
\end{align}
assuming $A_\mu = 0$,
or
\begin{align}
 \begin{pmatrix}
  r_1^2 \\ r_2^2 \\ r_3^2
 \end{pmatrix}
 = \frac 1{8m^2} \begin{pmatrix}
       -3\omega_1^2 +\omega_2^2 +\omega_3^2 \\
       \omega_1^2 -3\omega_2^2+\omega_3^2\\
        \omega_1^2 +\omega_2^2-3\omega_3^2
            \end{pmatrix}
\label{ri2-omega}
\end{align}
Note that these equations are independent of the representation $\pi_\mu$, and
have a large space of solutions as long as the rhs of \eq{ri2-omega} is positive.
The most obvious solution is obtained for coinciding
(time-like) frequencies\footnote{The sign of $\omega_i$ 
(forward or backward in time) is undetermined here, but will be fixed below.} 
$\omega_1^2=\omega_2^2 = \omega_3^2 = \omega^2$ and $r_i^2 = -\frac 1{8m^2} \omega^2$.
The scale of the background is thus set by the rotation frequencies $\omega_i$.

However, the assumption $A_\mu=0$ is consistent with Yang-Mills equations $D^\mu F_{\mu\nu} = \cJ_\nu$
only if the $\msu(N)$ gauge current $\cJ_\mu$ vanishes; this is not automatic 
for rotating backgrounds. We obtain
\begin{align}
 \cJ_{\mu} &= -i [\Phi_\a, D_\mu\Phi^\a]  
  = -4m^2 \sum_i r_i^2 \omega_{i\mu} H_i 
  \qquad \in \msu(N) \ ,
\end{align}
and therefore
\begin{align}
  \cJ_{\mu} &= 0 \qquad \mbox{if and only if} \quad  
    \sum_i r_i^2 \omega_{i\mu} \a_i  = 0 \ 
 \label{vanishing-J-cond}
\end{align}
or equivalently
\begin{align}
 r_1^2 \omega_{1\mu} =  r_2^2 \omega_{2\mu} = - r_3^2 \omega_{3\mu} \ .  
  \label{vanishing-J-cond-2}
\end{align}
This means that the three $U(1)_i$ R-currents $J_{\mu}^i$ \eq{J-i-explicit} must coincide. 
It turns out that taking into account this constraint,
 \eq{ri2-omega} admits only one solution,
\begin{align}
 r_1^2 &= r_2^2 = r_3^2 = -\frac 1{8m^2} \omega^2, \qquad \omega = \omega_1=\omega_2 = - \omega_3 \ 
 \label{allequal-constraint}
\end{align}
so that there is a single scale set by the time-like vector $\omega$.

Therefore we obtained rotating brane solutions of $\cN=4$ SYM, and 
similarly one obtains solutions of the IKKT matrix model on the quantum plane $\R^4_\theta$. 
Even though the above  analysis leaves only one solution \eq{allequal-constraint},
we will explore also the case of different $\omega_i$ in the following. 
This may arise in more complicated situations as discussed in section \ref{sec:interact-Higgs},
or possibly upon taking into account quantum effects which would modify the classical
equations \eq{ri2-omega}.

\section{Excitation modes on rotating branes}
\label{sec:fluctuations}

Consider internal fluctuations around this rotating brane background 
corresponding to 4-dimensional scalar fields,
\begin{align}
X_\a = \bar X_\a + \phi_\a   
\end{align} 
or more explicitly
\begin{align}
 X_i^\pm = r_ie^{\pm i\omega_i x}\, T_i^\pm  + \phi_i^\pm \ .
\end{align} 
Imposing the gauge condition $f=0$ \eq{gauge-fixing}, 
the action \eq{fluctuation-potential-complex} leads to the equation of motion\footnote{One may also
add a (positive) gauge-fixing  term $f^2$ to the potential as in \cite{Steinacker:2014lma}, 
which removes the unphysical pure gauge modes from the massless spectrum. This  
simplifies the analysis of the vector fluctuations. } 
for the fluctuations 
\begin{align}
 \big(\frac{\Box_4}{m^2} + \Box_X + 2 \slashed{D}_{ad}\big)\phi = 0, 
 \label{eom-fluct}
\end{align}
in terms of  the  ``adjoint'' Dirac operators
\begin{align}
 (\slashed{D}_{ad} \phi)_{\a} 
 &= [[\bar X_\a,\bar X^\b],\phi_\b] = \slashed{D}_{\rm mix} + \slashed{D}_{\rm diag} , \nn\\[1ex]
 (\slashed{D}_{\rm mix}\phi)_\a &=  \sum_{\b\neq \a}[[\bar X_\a,\bar X^\b],\phi_\b]   \nn\\
 (\slashed{D}_{\rm diag}\phi)_\a &= 2 [[\bar X_\a,\bar X_{-\a}],\phi_\a] 
\end{align}
with $X^\a \equiv 2 X_{-\a}$. Here
$\slashed{D}_{\rm mix}$ mixes the polarizations $\a$ (corresponding to roots of $\msu(3)$) 
and  is time-dependent, 
while $\slashed{D}_{\rm diag}$ preserves the polarizations and is static.
Using the time-dependent commutation relations for the background (cf.  \eq{Cartan-Weyl})
\begin{align}
 [\bar X_\a,\bar X_\b] &= e^{i(\omega_\a+\omega_\b-\omega_\g)x} r_\a r_\b r_\g^{-1} c_{\a\b}^\g \bar X_\g 
 + c_{\a\b}^i r_i H_i  \ ,
\end{align}
these are explicitly
\begin{align}
 (\slashed{D}_{\rm mix}\phi)_\a &=  \pm 2 \sum\limits_{\b \neq \a} e^{i(\omega_a-\omega_\b-\omega_{\a-\b})x} r_\a r_\b r_\g^{-1}  [X_{\a-\b}, \phi_\b]  \nn\\
 (\slashed{D}_{\rm diag}\phi)_\a &= 2 r_\a [H_\a, \phi_\a]  \qquad \mbox{(no sum)} .
 \label{D-ad-roots}
\end{align}
Only terms with $\a-\b\in\cI$ contribute in the first line.
We note that 
\begin{align}
 \tau \slashed{D}_{\rm mix} = -\slashed{D}_{\rm mix} \tau \ .
 \label{tau-D-anticomm}
\end{align}
where $\tau = \diag(i,i,-i)$ is the generator of 
$U(1)\subset SO(6)$ corresponding to the simultaneous rotations 
of the $\phi_\a$ along the $\a_1,\a_2$ and $-\a_3$ directions (cf. figure \ref{fig:spin-states-roots}).
The equations of motion  separate into time-independent and time-dependent terms,
\begin{align}
 (\frac{\Box_4}{m^2} + \Box_X  + 2\slashed{D}_{\rm diag} \big) \phi 
 + 2\slashed{D}_{\rm mix}  \phi =0 .
 \label{eom-separate}
\end{align}
Let us spell out these equations in
the complex root basis, starting with \eq{eom-fluct}
\begin{align} 
 (\frac{\Box_4}{m^2} + \Box_X) \phi_i^+ + 4 \sum_j r_i r_j e^{i\omega_ix}
   \Big(e^{-i\omega_jx} [[T_i^+,T_j^-],\phi_j^+]
  +  e^{i\omega_jx} [[T_i^+,T_j^+],\phi_j^-]\Big) &=0 
 \end{align} 
along with the 3 conjugate equations. Using the $\msu(3)$ relations
$[T_i^-,T_j^+] = \d_i^j H_i$ 
for $i=1,2$,  $[T_1^+,T_3^-] = - T_2^-$ and $[T_2^+,T_3^-] =  T_1^-$,
the explicit form of \eq{eom-separate} is
\begin{align}
  \big(\frac{\Box_4}{m^2} + \Box_X + 4 r_1^2 [H_1,.]\big) \phi_1^+ 
 + 4 r_1e^{i\omega_1x}\Big(r_2 e^{i\omega_2x} [T_3^+,\phi_2^-] - r_3 e^{-i\omega_3x} [T_2^-,\phi_3^+]\Big) &=0 \nn\\
 \big(\frac{\Box_4}{m^2} + \Box_X + 4 r_2^2 [H_2,.]\big) \phi_2^+ 
 + 4 r_2 e^{i\omega_2x}\Big( r_3 e^{-i\omega_3x} [T_1^-,\phi_3^+]
  - r_1e^{i\omega_1x} [T_3^+,\phi_1^-]\Big) &=0 \nn\\
  \big(\frac{\Box_4}{m^2} + \Box_X + 4 r_3^2 [H_3,.]\big) \phi_3^+ 
  + 4 r_3 e^{i\omega_3x}\Big(r_2 e^{-i\omega_2x} [T_1^+,\phi_2^+] -r_1 e^{-i\omega_1x} [T_2^+,\phi_1^+]  \Big) &=0 
\label{eom-phi-explicit}
\end{align}
as well as the conjugate relations.
Note that the first terms are time-independent and diagonal in the 
polarization, while the second terms are time-dependent and mix the polarizations.
It turns out that the solutions to these equations fall into two  classes:
First, there is a set of zero modes (or ``would-be zero modes'' as explained below),
which are oblivious to the background rotation
and have standard dispersion relation.
All other modes couple to the background rotation, and accordingly have deformed dispersion relations.
This could be physically acceptable if they are massive in some generalized sense.

\subsection{Zero modes}
\label{sec:zero-modes}

There is a special set of modes $\phi^{(0)}$ characterized by 
$\slashed{D}_{\rm mix}\phi^{(0)}=0$, which are thereby oblivious to the 
explicit $x^\mu$-dependence of the background, and which are zero modes if all $r_i$ coincide.
They will be denoted as zero modes, or ``would-be zero modes'' for reasons explained below.
In view of
\eq{D-ad-roots} or \eq{eom-phi-explicit}, these modes must satisfy   
\begin{align}
T_{\a-\b} \triangleright \phi_\b^{(0)}:=  [T_{\a-\b}, \phi_\b^{(0)}] =0 
   \qquad \forall \a, \b \ \mbox{with} \ \a-\b\in\cI .
 \label{time-indep}
\end{align}
or equivalently
\begin{align}
 T_{\a} \triangleright \phi_\b^{(0)} = [T_{\a}, \phi_\b^{(0)}] = 0 \quad \mbox{for} \quad \a+\b\in\cI 
 \label{zeromode-cond-2}
\end{align}
Here $\triangleright$ denotes the $\msu(3)$ action on
\begin{align}
 \phi_\a \in End(\cH) = \oplus V_\L .
 \label{functions-decomp}
\end{align}
It is easy to find all solutions of \eq{time-indep}: 
These are two equations for each $\phi_\b$, which state that 
$\phi_\b$ is an extremal weight\footnote{A weight $\L$ in a (finite-dimensional) representation $V$ is called extremal if 
it is related by some element $w\in\cW$ to a highest weight of $V$.}
vector in $V_\L$ with weight
in the Weyl chamber opposite to $\b$.
For example, take $\a=-\a_3$.
Then $T_1^+ \triangleright\phi_{-\a_3}^{(0)} = 0 = T_2^+ \triangleright\phi_{-\a_3}^{(0)}$ 
is tantamount to the statement that
$\phi_{-\a_3}$ is a highest weight vector in $V_\L$,
\begin{align}
\phi_{-\a_3}^{(0)}  = |\L,\L\rangle .
 \label{zeromodes-higgs-explicit} 
\end{align}
Analogous modes are obtained by acting with the 
Weyl group\footnote{We recall that the Weyl group $\cW$ -- 
or more precisely a certain covering $\widetilde\cW$ in the
braid group -- can be viewed as a discrete subgroup of $SU(3)$, and therefore acts on any finite-dimensional 
representation of $\msu(3)$.}
$\cW$ on $\phi_{-\a_3}^{(0)}$, denoted as
\begin{align}
 \phi_{-\a}^{(0)} =  |w_\a\L,\L\rangle, \qquad i = 1,2,3 
 \label{zeromodes-explicit}
\end{align}
where $w_\a$ maps the fundamental Weyl chamber into that of $\a$, i.e.
$w_{\a}\cdot \a_3 = \a$. The gauge-fixing condition $f=0$ is satisfied.
They all satisfy \eq{time-indep} and 
\begin{align}
 \slashed{D}_{\rm mix}\phi_\a^{(0)} =0 , 
 \label{D6-vanishing}
\end{align}
and thus are oblivious to the 
background rotation.
Particular examples\footnote{For squashed $\C P^2_N$, this exhausts all zero modes in $End(\cH_\mu)$.} 
of such modes are given by 
\begin{align}
 \phi_i^{\pm(0)} = (T_i^\mp)^{n} ,
 \label{zeromodes-examples}
\end{align}
for $n$ below some cutoff.
As a consequence, the equations of motion reduce to
\begin{align}
(\frac{\Box_4}{m^2} + \Box_X + 2\slashed{D}_{\rm diag}\big)\phi^{(0)} =0 .
\label{eom-reduced}
\end{align}
We evaluate $\Box_X$ explicitly for 
$\phi_{-\a_3}^{(0)} = |\L,\L\rangle$: 
\begin{align}
 \frac 12 \Box_X |\L,\L\rangle &= \sum_i [X_i^+,[X_i^-,.]] |\L,\L\rangle  
   = \sum_i r_i^2\, T_i^+ T_i^- |\L,\L\rangle  \nn\\
   &=  \sum_i r_i^2\, [T_i^+, T_i^-] |\L,\L\rangle  
   = \sum_i r_i^2 H_i |\L,\L\rangle \nn\\
   &= \big(r_1^2\a_1 + r_2^2 \a_2 + r_3^2 \a_3,\L\big) |\L,\L\rangle .
\end{align}
Furthermore,
\begin{align}
\slashed{D}_{\rm diag}\phi^{(0)}_{-\a_3}  &= 2r_3^2 [[T_{-3},T_{3}],\phi_{-\a_3}^{(0)}] 
  =  - 2r_3^2 [H_3,\phi_{\a}^{(0)}]  \nn\\
  &=  -2 r_3^2\, (\a_3,\L) \phi_{-\a_3}^{(0)}
\end{align}
and  the eigenvalue on $\phi^{(0)}_{\a_3}$ is the same.
Therefore we have obtained the $\a=-\a_3$ case of
\begin{align}
(\Box_X + 2\slashed{D}_{\rm diag})\phi^{(0)}_{\a}  = M_{\a}^\phi \phi^{(0)}_{\a},  
\label{M-phi-EV}
\end{align}
with 
\begin{align}
 M_{-\a_3}^\phi &= 2\big(r_1^2\a_1 + r_2^2 \a_2 - r_3^2 \a_3,\L\big) .
\label{Higgs-mass}
\end{align}
Similarly for $\phi_{\a_3}^{(0)} = |\Omega\L,\L\rangle$ where $\Omega\L$ is the extremal weight in the opposite 
Weyl chamber to $\L$, we obtain
\begin{align}
 M_{\a_3}^\phi &= 2\big(-r_1^2\a_1 - r_2^2 \a_2 + r_3^2 \a_3,\Omega\L\big)
\end{align}
and $ M_{-\a_3}^\phi =  M_{\a_3}^\phi$ if $\Omega\L = -\L$.
This clearly vanishes if all $r_i$ or $\omega_i^2$ coincide.
In general, the mass  of $\phi_{\a}^{(0)}$ is given by
\begin{align}
  M_{\a}^\phi &= 2\big(\s_1 r_1^2\a_1 + \s_2r_2^2 \a_2 + \s_3r_3^2 \a_3,\L'\big) \nn\\
   &= 0 \qquad \mbox{if}\quad  r_1 = r_2 = r_3 \ 
\end{align}
where $\s_i$ is the sign of the Weyl group element relating $\a_i$ with 
the extremal weight $\L'$ (which is in the Weyl chamber opposite to $\a$).
This becomes more transparent in terms of the over-complete weight labels 
\begin{align}
 n_1 = (\L,\a_1), \quad n_2 = (\L,\a_2), \quad n_3 = (\L,-\a_3) = -(n_1+n_2) .
\end{align}
Then the  eigenvalues can be written as
\begin{align}
\begin{pmatrix}
 M_{\a_3}^\phi \\ M_{-\a_2}^\phi \\ M_{-\a_1}^\phi
\end{pmatrix}
 = 2\begin{pmatrix}
    n_1 & n_2 & n_3\\
    n_2 & n_3 & n_1\\
    n_3 & n_1 & n_2
   \end{pmatrix}
\begin{pmatrix}
 r_1^2 \\ r_2^2 \\ r_3^2 
\end{pmatrix}  
 \label{M-phi-results}
\end{align}
and similarly for the opposite roots.
While the individual masses can have either sign, they satisfy the sum rule
\begin{align}
 M_{-\a_1} ^\phi + M_{-\a_2}^\phi + M_{\a_3}^\phi  &=0 
 \label{sum-rule}
\end{align}
and similarly for the opposite roots.
For $r_1=r_2=r_3$, we recover
$M_i^\phi=0$, and  all $\phi^{(0)}_{\a}$ are exact zero modes.  
Conversely, we show in appendix \ref{app:potential} that for $r_i=r$, all zero modes of the 
time-independent potential $(\Box_X + 2\slashed{D}_{\rm diag})$ are indeed given by the 
above  modes $\phi_\a^{(0)}$, which in turn are zero modes of $\slashed{D}_{\rm mix}$.
Therefore we have properly identified the regular modes
which obey standard Lorentz-invariant kinematics. 
Since these $\phi^{(0)}_{\a}$ modes decouple from the rotation also 
for different $r_i$, we will denote them as would-be zero modes, 
as they may in general acquire non-vanishing masses $M_\a^2$.

Finally we need to take into account the reality condition 
$(\phi_i^+)^\dagger = \phi_i^-$, mapping the extremal weights $\L$ to $-\L$.
For real representations $V_\L$, the reality condition
relates the two opposite extremal modes, so that 
\begin{align}
 \big(\phi_i^{+(0)}\big)^\dagger = |-w_i^-\L,\L\rangle = |w_i^+\L,\L\rangle = \phi_i^{-(0)}.
\end{align}
For complex representation, which arise in particular for links between different branes, 
the reality condition relates zero modes with opposite extremal weights in the conjugate representations
$V_\L$ and $V_\L^* = V_{-\Omega\L}$.

To summarize, we have found 6 real or 3 complex  zero modes
$\phi_i^{\pm(0)}(k) e^{i k x}$ on $\R^4$, corresponding to the 
extremal weights $\cW \L$ in the decomposition $End(\cH) = \oplus V_\L$.
As discussed in \cite{Steinacker:2014lma}, they 
can be interpreted as oriented strings linking the 6 
coincident sheets\footnote{assuming that $\mu$ has no stabilizer in $\cW$;
for  squashed $\C P^2_N$ they link the 3 intersecting sheets \cite{Steinacker:2014lma}.} 
of $\cC_N[\mu]$ with opposite flux. This is most obvious for the maximal weight state 
$|\L,\L\rangle = |\mu\rangle\langle\Omega \mu|$ in $End(\cH_\mu)$,
since the $|\mu\rangle$ can be interpreted 
as coherent state  on a sheet of $\cC_N[\mu]$ localized at the origin.
Then  conjugate modes correspond to reversed links, which are identified by the hermiticity
constraint.
For negative mass $M_i^2<0$, the mode will be stabilized by the quartic terms in the interaction,
thus leading to a non-trivial vacuum expectation value $\langle\phi_i^{(0)}\rangle \neq 0$.
This should introduce a low-energy scale into the massless sector, as discussed below.

\subsection{The deformed or mixing sector}

Now we want to find the general solutions of the equation of motion for the fluctuations 
\begin{align}
 \big(\Box_4 + m^2(\Box_X + 2 \slashed{D}_{\rm diag})\big) \phi \ + \  2m^2\slashed{D}_{\rm mix}(x) \phi = 0
\label{scalar-eom-heavy}
\end{align}
where $\Box_4 = -D^\mu D_\mu$.
The explicit $x^\mu$-dependence of the mixing term is indicated by writing
$\slashed{D}_{\rm mix}(x)$. 
Writing the rotating background  \eq{rotating-brane} in the form
\begin{align}
 X_\a = r_\a (e^{\omega_i x\tau_i} T)_\a
\end{align}
(sum over $i$ is understood)
where $\tau_i$ is the generator of the
$U(1)_i\subset SO(6)_R$  rotation, we have
\begin{align}
 \slashed{D}_{\rm mix}(x) = e^{\omega_i x \tau_i}\,\slashed{\obar D}_{\rm mix}\,e^{-\omega_i x \tau_i}
\end{align}
where $\slashed{\obar D}_{\rm mix}$ is  independent of $x$.
Accordingly, we make an ansatz 
 \begin{align}
 \phi^{(n)}_k(x) &= e^{(i k + \omega_{i}\t_i)x }\, \obar \phi^{(n)}  + h.c.
 =  e^{i p x} \obar \phi^{(n)} + h.c.,  \qquad p = k - i\omega_{i} \t_i
 \label{phi-n-transl}
\end{align}
for the excitation modes around the rotating background.
Noting that $\t = \sum_i \t_i$ satisfies $\t^2 = -\one$,
equation \eq{scalar-eom-heavy} turns into the algebraic eigenvalue problem
\begin{align}
  (p^2 + m^2 \cO_V)\obar\phi^{(n)} = 0
\label{scalar-eom-heavy-EV}
\end{align}
where 
\begin{align}
 \cO_V =  \Box_X +  2 \slashed{D}_{\rm diag} + 2\slashed{\obar D}_{\rm mix}\ .
 \label{OV-def}
\end{align}
Implementing \eq{scalar-eom-heavy} on a computer and 
restricting ourselves to $k_\mu = (k_0,\vec 0) \sim \omega_\mu$ modes, we obtain  the following spectrum 
 on a minimal brane  $\cC[(1,0)]$
\begin{align}
k_0^2 \in \{15\times 0, 12 \times 8, 6\times 24, 6\times 32, 2\times 36, 48, 6\times 4 (5 + 3 \sqrt{3}), 
6\times 4 (5 - 3 \sqrt{3})\} .
\label{spec-rotating}
\end{align}
Unfortunately there is a sextet of modes 
with imaginary frequency $k_0^2 = 4 (5 - 3 \sqrt{3}) \approx -0.78$, which means 
that the background is classically unstable.
However since this value is rather small, one may hope that these modes 
are stabilized by quantum effects\footnote{In particular, one expects an attractive interaction 
of such branes at one loop due to the relation with supergravity, 
which should help to stabilize them.} for moderate coupling.
More detailed examination shows that these unstable modes arise from (mixtures of) the 
massive modes $6 \times (12 \pm 4 \sqrt{3})$ on a non-rotating background \eq{spec-OV}. 
Hence a mild modification of these 
-- e.g. through quantum corrections -- might suffice to stabilize the background.


Let us  discuss the kinematics of these  modes in some more detail, ignoring the 
unstable modes for now. It is useful to observe that our time-dependent background admits a modified  
translation invariance $\cT^4 \cong \R^4$, generated by
\begin{align}
 V_\mu = \del_\mu - \omega_{i\mu} \t_i \ .
\end{align}
Thus the above ansatz \eq{phi-n-transl}  states that the  $\phi^{(n)}_k(x)$ are 
unitary\footnote{except for the unstable modes, which are non-unitary representations.} irreps 
of these generalized translations,
 \begin{align}
 V_\mu \phi^{(n)}_k(x) &= i k \phi^{(n)}_k(x) \ .
\end{align}
$\cT^4$ is also respected in the interactions, because the $U(1)_i$ are part of the $SO(6)_R$ symmetry.
Thus the generalized momentum $V_\mu$ resp. $k_\mu$ is conserved rather than
the usual momentum $p_\mu$ associated to $\del_\mu$, and Poincare invariance is typically 
broken for the heavy sector.
However, the zero modes $\phi^{(0)}_k(x)$ can be chosen as eigenfunctions 
not only of $V_\mu$ but also of the $\tau_i$ and $\del_\mu$, with 
\begin{align}
  \del_\mu \phi^{(0)\pm i}_k(x) = i (k_\mu \pm \omega_{i\mu})\,\phi^{(0)\pm i}_k(x) , \qquad i=1,2,3
\end{align}
This is obvious due to their explicit form \eq{zeromodes-explicit}; these turn into 
the 12 modes with $k_0^2 = 8$ in \eq{spec-rotating}. 
In terms of the  usual momentum $p_\mu$, they have an undeformed massless dispersion relation.
Moreover, we will see that $p_\mu$ is conserved even in the interactions of 
the  zero modes $\phi^{(0)}_k$. 
Therefore this zero mode sector could lead to standard Poincare-invariant physics, 
assuming the resonance modes get stabilized.
All other modes have a modified dispersion relation, which would be acceptable if 
they become heavy due to quantum effects. 

%

There is an interesting relation with the non-rotating case $\omega_i = 0$,
which provides also some checks for these numerical results.
Observe that $\cO_V$ is unitarily equivalent
to 
\begin{align}
 \t \cO_V \t^{-1} =  \Box_X + 2\slashed{D}_{\rm diag} - 2\slashed{D}_{mix}  \ ,
\end{align}
using $\t {D}_{mix} \t^{-1} = - {D}_{mix}$ \eq{tau-D-anticomm}. The operator on the rhs
gives precisely the potential  (B.7) considered in \cite{Steinacker:2014lma} 
in the presence of a cubic flux term\footnote{This is true only 
for the form \eq{scalar-eom-heavy}, i.e. after dropping the 
``pure gauge'' term $[X^\a,[X_\b,.]]$ in \eq{fluctuation-potential-complex}.} , 
which was shown to be positive semi-definite for all representations.
Now for $\omega_i=0$, we obtain the following spectrum of $\cO_V$
for the minimal brane  $\cC[(1,0)]$:
\begin{align}
 {\rm spec} \cO_V = \{18\times 0, 6 \times (12 - 4 \sqrt{3}),13\times 8, 2\times 12, 6\times 16,
 6\times (12 + 4 \sqrt{3}), 2\times 20, 1 \times 24\}
 \label{spec-OV}
\end{align}
This reproduces the spectrum of 
the potential (B.7) in \cite{Steinacker:2014lma}, which has 6 exceptional zero modes 
in addition to the 12 regular zero modes.
This provides, incidentally, a good  starting point to compute 
the one-loop effective potential for squashed branes in $\cN=4$ SYM:
although the static branes are not  classical solutions, 
one can still ask the question whether they might be stabilized by quantum effects. 
We hope to pursue this question elsewhere.

\subsection{Fermions}
\label{sec:fermions}

\begin{figure}
\begin{center}
 \includegraphics[width=0.45\textwidth]{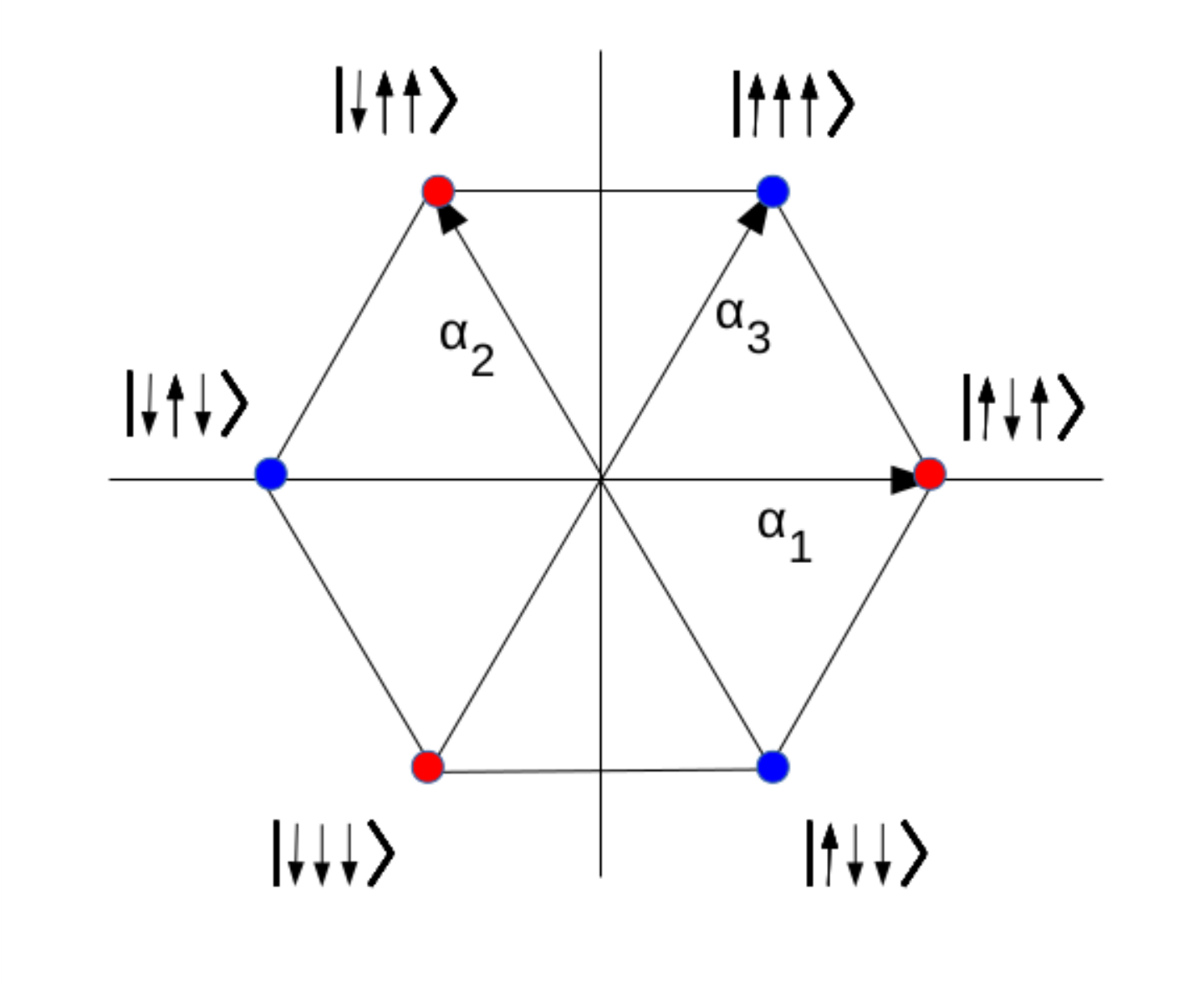}
 \end{center}
 \caption{Fermionic zero modes in weight space with roots $\a_i$,  and chirality 
 sectors ${\color{red} C_L}$, ${\color{blue} C_R}$  indicated by color.}
 \label{fig:spin-states-roots}
\end{figure}

The spectrum of 4-dimensional fermions and their masses is governed by the Dirac operator  on 
squashed $\cC_\cN[\mu]$. In the time-dependent background, it can be written as 
\begin{align}
\slashed{D}_{(6)} \Psi = \sum_{a\in\cI} \Delta_a [X_a,\Psi] 
 &= 2\sum_{i=1}^3 \Big(\Delta^-_i [X_i^+,.] + \Delta_i^+ [X_i^-,.]\Big) \nn\\
 &= 2\sum_{i=1}^3 r_i\Big( \Delta^-_i e^{i\omega_i x} ad_{T_i^+} + \Delta_i^+ e^{-i\omega_i x} ad_{T_i^-}\Big)
 \label{Dirac-ladder}
\end{align}
where the spinorial ladder operators
\begin{align}
2\Delta^-_1 &= \Delta_4 - i \Delta_5, \qquad 2\Delta_1^+ =  \Delta_4 + i \Delta_5,   \nn\\
2\Delta^-_2 &= \Delta_6 + i \Delta_7, \qquad 2\Delta_2^+ =  \Delta_6 - i \Delta_7,   \nn\\
2\Delta^-_3 &= \Delta_1 - i \Delta_2, \qquad 2\Delta_3^+ =  \Delta_1 + i \Delta_2,   
\end{align}
 satisfy
\begin{align}
\qquad  \{\Delta^-_i,\Delta_j^+\} = \d_{ij}. 
\end{align}
In particular, the partial chirality operator on $\R^2_i$ is  given by 
\bea 
\chi_i &=&  -2(\Delta_i^\dagger \Delta^-_i-\frac 12), 
\label{sub-chirality}
\eea
acting on the spin-$\frac 12$ irreducible representation. 
Using the form \eq{Dirac-ladder}, we can easily find the zero modes as in \cite{Steinacker:2014lma}: 
let $Y^\L_\L$ be the highest weight vector of $\cH_\L \subset End(\cH_\mu)$. Then 
\begin{align}
\slashed{D}_{(6)} \Psi_{\L} &= 0  \qquad \mbox{for} \qquad \Psi_{\L} 
= |\uparrow\uparrow\uparrow\rangle \, Y^\L_\L \ .
\end{align}
This follows immediately from the decomposition \eq{Dirac-ladder} of the Dirac operator,
noting that
\begin{align}
  \Delta_i^+|\uparrow\uparrow\uparrow\rangle &= 0 .
\end{align}
Analogous zero modes $\slashed{D}_{(6)} \Psi_{\L'} = 0 $ are obtained for any 
extremal weight $\L' \in\cW\L$ in $V_\L$:
\begin{align}
  \Psi_{\L'} &= |s_1,s_2,s_3\rangle  Y^\L_{\L'} 
 \label{fermion-zeromodes}
\end{align}
for suitable spin states $|s_1,s_2,s_3\rangle$.
They can be obtained successively by applying a (generalized) Weyl reflection 
relating extremal weights $\L'$  in adjacent Weyl chambers as follows:
\begin{align}
 \Psi_{w_i\L'} = \tilde \omega_i \cdot\Psi_{\L'} 
 :=  (\omega_i |\uparrow\uparrow\uparrow\rangle)\, (w_i\cdot Y^\L_{\L'}) \ ,
\end{align}
{\em provided} $\L'$ and $w_i\L'$ are extremal weights of $V_\L$ in adjacent Weyl chambers.
Here $\omega_i$ implements the Weyl reflection $w_i$ on the internal spinor space $(\C^2)^{\otimes 3}$
associated to the $\Delta_{1,...,6}$ as follows:
\begin{align}
 \omega_i \Sigma_{(i)}  \omega_i^{-1}  &= -\Sigma_{(i)} , \qquad i = 1,2,3 \nn\\
 \omega_i \Sigma_{(j)}  \omega_i^{-1}  &= \Sigma_{(j)} , \qquad j\neq i \nn\\
 \Sigma_{(i)} &= \frac 12[\Delta^-_i,\Delta_i^+] 
  = \frac 12\chi_i \ .
 \label{Weyl-spinors}
\end{align}
For example if $\L = \a_3$, then $\Psi_{w_3\L} = \tilde\omega_2 \cdot \tilde\omega_3\cdot \tilde\omega_1\cdot\Psi_\L$;
note that the $\tilde\omega_i$ do not satisfy the Weyl group relations, but this is not a problem here.
These spin states are visualized in figure \ref{fig:spin-states-roots}, and 
fall into chirality classes  $C_L$ and $C_R$ with
 well-defined internal chirality 
\begin{align} 
 \chi  \Psi_{w\L} &= (-1)^{|w|}  \Psi_{w\L} \ .
 \label{chirality}
\end{align} 
These zero modes have definite chirality on $\R^4$, 
because $\Psi$ is subject to the Majorana-Weyl condition 
$\Psi^C = \Psi = \Gamma\Psi$.
Moreover, it was shown in \cite{Steinacker:2014lma} that the extremal 
modes $\Psi_\L$ and $\Psi_{-\L}$ are related by
(internal) charge conjugation and have opposite chirality,
\begin{align}
 C^{(6)} \Psi_\L^* = \Psi_{-\L} .
\end{align}
As long as  $\L$ has no stabilizer in $\cW$, we can 
denote them as $\Psi_i^\pm$, identifying $\L$ with the root $\a_i^\pm$ in the same Weyl chamber.
Taking into account the  Majorana-Weyl condition, 
this implies that the 
corresponding solutions of the full Dirac operator have the form \cite{Steinacker:2014lma}
\begin{align}
 \Psi_i(x) = \Psi_i^+ \otimes \psi^i_+(x) + \Psi_i^- \otimes \psi^i_-(x),
 \label{MW-spinor-full}
\end{align}
where the four-dimensional spinors $\psi^i_\pm$ satisfy
\begin{align}
 \Di_{(4)} \psi_\pm^i(x) & = 0, &
 \gamma_5 \psi_\pm^i(x) & = \pm \psi_\pm^i(x), &
 (\psi_\pm^i(x))^C & = \psi^i_\mp(x).
 \label{MW-explicit}
\end{align}
This means that the $\psi_\pm^i$ are not independent,
as $\psi_+^i(x)$ determines $\psi_-^i(x)$.
We can  expand the general solution in terms of 
plane wave Weyl spinors $\psi^\pm_{i;k}(x)$ on $\R^4$ with  momentum $k$, 
\begin{align}
 \Psi_i(x) = \int \frac{d^3 k}{\omega_k} \big(\psi_{i;k}^+(x)  \Psi^i_+
                                      + \psi^{-}_{i;k}(x) \Psi^i_- \big) , \qquad i= 1,2,3 .
\end{align} 
This can be viewed either in terms of three 4-dimensional Majorana spinors
$\psi_{i}^+  + \psi_{i}^-$,
or three Weyl spinors $\psi^{+}_i$ in 4D.

To summarize, the fermionic zero modes are in one-to-one 
correspondence with
extremal weights $\L$, with $\L$ related to $-\L$ by charge conjugation
precisely as the scalar zero modes.
Hence these modes can be succinctly collected in terms of chiral $\cN=1$ multiplets
labeled by positive extremal $\L$.
Moreover, the chirality \eq{chirality} of the fermionic zero modes is distinguished 
by their charges under the noncommutative gauge fields on 
squashed $\cC_N^6[\mu]$ linked by $\Psi_{w\L}$. 
This can be seen in figure \ref{fig:spin-states-roots} for the gauge field modes $A_\mu^{(3,8)}(x) T_{3,8}$ 
corresponding to the Cartan generators of $\msu(3)$, which are among the lowest non-trivial gauge field modes 
according to section \ref{sec:gauge-bosons}.
It is even more obvious for the gauge fields corresponding to $\Theta$ \eq{theta-def},
as discussed in the next section.
In other words, different chiralities have different gauge couplings, which is a signature of
a chiral model.
Of course the total index $\tr \gamma_5$ of the zero modes vanishes and the model 
is guaranteed to te anomaly free,  suggesting some left-right symmetric model. 
However, more interesting behavior is possible in suitable brane configurations,
as explained in section \ref{sec:standard-model}.

We have also seen that in the case $r_i\neq r_j$, some of the scalar fields acquire a negative mass,
which may lead to  non-trivial vacuum expectation values and mass terms for these fermions.

If $\L$ has non-trivial stabilizer $w_i\in\cW$, the above analysis goes through, but
with two fermionic zero modes with opposite chirality for each (positive) extremal weight $\L$.
There is a similar doubling for the scalar zero modes, since then $\L$
is in the Weyl chamber opposite to two roots. Therefore the fermionic and  bosonic zero modes
can still be grouped into  chiral supermultiplets with opposite chirality.
Finally, the $\L=0$ modes correspond to a $\cN=4$ 
supermultiplet of trivial $\L=0$ modes.

\subsection{Gauge bosons}
\label{sec:gauge-bosons}

The 4-dimensional gauge bosons $A_\mu(x)$ decompose into eigenmodes of the 
scalar Laplacian $\Box_X$.
Since the scalar Laplacian is time-independent, 
all these modes are governed by a standard 4-dimensional
kinematics, in contrast to the heavy Higgs modes discussed above.
It is easy to compute its spectrum for equal $r_i=r$, observing that
 $\Box_X = r^2(C_2 - [H_3,[H_3,.]] - [H_8,[H_8,.]])$ where $C_2$ is the quadratic Casimir
of $SU(3)$.
This gives 
\begin{align}
  \Box_X|M,\L\rangle 
  &= 2r^2 \l_{\L,M} |M,\L\rangle  , \qquad  
 \l_{\L,M} =  (\L,\L +2\rho) - (M,M)
  \label{Box-X-mass}
\end{align}
so that the $|M,\L\rangle$ states are indeed eigenstates of $\Box_X$.
As a check, we recover 
\begin{align}
 \Box_X X_i^\pm = 2r^2 (\L,2\rho) X_i^\pm = 8 r^2 X_i^\pm
\end{align}
since $X_i^\pm \in |\pm \L,\L\rangle$ with $\L=\rho=\a_3$ and $(\a_3,\a_3) =2$.
As shown in  \cite{Steinacker:2014lma}, it follows that the only zero modes are those with $M=\L=0$,
while the lowest non-vanishing eigenvalues arise for $M= w \L$ 
with eigenvalues $4(\L,\rho) \geq 8r^2$.
Via the  Higgs effect elaborated in \eq{Higgs-gaugebosons-action}, these modes acquire a mass
\begin{align}
 m_A^2 = 2 m^2 r^2 \l_{\L,M} \ = \ -\frac 14 \omega^2 \l_{\L,M} 
 \label{gauge-boson-mass}
\end{align}
using \eq{ri2-omega} and 
assuming that $\omega_i^2 = \omega^2$.

We will see in section \ref{sec:higgs-gauge} that only the 
the $M= 0$ gauge modes couple to the  currents arising from the low-energy sector
of fermions linking point branes with $\cC[\mu]$; these will be of primary interested
from the particle physics perspective.
The lowest such modes are the two weight zero modes 
in $\L=(1,1)$ given by the Cartan generators $H_3, H_8 \in \msu(3)$,
with mass $m_A^2 =  12 m^2 r^2$.
However let us focus on the next-lowest modes given by the $M=0$ modes in  $(0,3)$ and $(3,0)$,
which lead to particularly interesting  chiral gauge fields.
They arise in the algebra of functions on $\cC_N[\mu]$ as follows:
Consider the functions
\begin{align}
E_{ab} &=  \varepsilon^{(8)}_{abcdefgh} T^cT^dT^eT^f T^g T^h 
 = \frac 18\varepsilon^{(8)}_{abcdefgh}[T^c,T^d][T^e,T^f][T^g,T^h]  = - E_{ba}  \nn\\
 &= h_{abcde} T^cT^dT^e
 \label{Eab-def}
\end{align}
with $\msu(3)$ indices in $\{1,...,8\}$, which 
form a multiplet $(8)\wedge (8) = (3,0) \oplus  (0,3) \oplus  (1,1)$.
It is not hard to see that the $(1,1)$ components do not occur\footnote{
they could only be $\sim f^{ab}_c T^c$, however this is ruled out by 
$\tr(\varepsilon^{(8)}_{abcdefgh} T^a T^b T^cT^dT^eT^f T^g T^h) = 0$.},
so that the $E^{ab}$ decompose into $(3,0) \oplus  (0,3)$.
We focus on the two $M=0$ modes in  $(3,0)$ and $(0,3)$. 
Having weight zero means that 
they commute with the Cartan generators $H_\a$, and  diagonalize on the extremal
weight states in $End(\cH)$.
One hermitian combination of these is given by the 
orientation form $\Theta$, defined by
\begin{align}
 \Theta &:= iE_{3,8} = i\varepsilon^{(6)}_{abcdef} T^aT^bT^cT^dT^eT^f 
 = \frac i8\varepsilon^{(6)}_{abcdef}[T^a,T^b][T^c,T^d][T^e,T^f] \nn\\
  &\sim \frac 18\varepsilon^{(6)}_{abcdef}\{x^a,x^b\}\{x^c,x^d\}\{x^e,x^f\} 
   = {\rm Pf}\theta^{ab} 
  \label{theta-def}
\end{align}
with indices restricted to the roots.
Here $\theta^{ab} = \{x^a,x^b\}$ is the Poisson structure on $\cC[\mu]$
with coordinate functions $x^a$ arising form the matrices $T^a$ in the semi-classical limit.
This is invariant under the $\Z_3$ subgroup of Weyl rotations,  odd under reflections, and
reduces in the semi-classical limit to the (symplectic) orientation form ${\rm Pf}\theta^{ab}$
on the left- and right-handed sheets
$C_L$ and $C_R$  of the 6-dimensional branes $\cC_N[\mu]$ \cite{Steinacker:2014lma}.
Therefore $\Theta$ takes eigenvalues $\pm c$ on
the left- and right-handed maximal states $\Psi_{\L'}$, 
which  is a crucial ingredient of a chiral 
gauge theory\footnote{The other hermitian combination is given by any of the three forms
$\Xi := E_{\a_1,-\a_1} \sim E_{\a_2,-\a_2} \sim - E_{\a_3,-\a_3}$,
which is invariant under the full Weyl group
(since e.g. 
$w_1 E_{\a_1,-\a_1} = E_{\a_1,-\a_1}$.).
Numerical computations indicate that $\Xi$ vanishes on the extremal states of $\cH_\mu$,
at least for small $\mu$.}.

Finally for different $r_i$ , it is clear that the trivial zero modes $M=\L=0$ persist as above,
and the massive modes will get somewhat deformed, but 
preserving the qualitative features as long as $r_i \approx r$.
The $M=\L=0$ modes  become interesting in the case of
$n$ coincident branes, where the turn into nonabelian $U(n)$ gauge bosons.

\section{Interactions}

\subsection{Scalar self-interactions and nontrivial Higgs}
\label{sec:interact-Higgs}

Now consider the  interactions of the scalar zero modes discussed in section \ref{sec:zero-modes}.
The quartic and cubic self-interactions are easily obtained from \eq{full-potential},  
\begin{align}
 V_{\rm int}(\phi) &= \frac{m^4}{g^4}\tr \Big( X_\a\Box_\phi \phi^\a 
 - \frac 14 [\phi_\a,\phi_\b][\phi^\a,\phi^\b] \Big)
 \label{scalar-interaction}
\end{align}
The quartic term will stabilize the (would-be)  zero modes.
The cubic term coupling to $X_\a$ turns out to
 vanish identically for the zero modes.
This can be seen by writing it different ways
\begin{align} 
  \tr X^\a\Box_\phi \phi_\a 
 &=  \tr[X_\a,\phi_\b][\phi^\a,\phi^\b] \nn\\
  &=  -\tr \phi_\b [[\phi^\a,\phi^\b],X_\a] 
  =  \tr \phi_\b \Big([[\phi^\b,X_\a],\phi^\a]  + [[X_\a,\phi^\a],\phi^\b] \Big)\nn\\ 
  &=  -2\tr \phi_\b [[\phi_{-\b},X_\a],\phi^\a] 
  \label{V-phi-rewrite}
\end{align} 
using the Jacobi identity, $\phi^\b = 2\phi_{-\b}$, and the gauge-fixing condition 
\begin{align}
 [X^\a,\phi_\a] &= 0 
 \label{decoupl-sum}
\end{align}
which holds for the zero modes discussed in section \ref{sec:zero-modes}.
Now recall that
 $[X_\a,\phi_\b]=0$ if  $\a+\b \in \cI$ or $\a+\b=0$,
due to \eq{zeromode-cond-2}; this amounts to the extremal weight property
of the zero modes.
Similarly  $[X_{\a},\phi_{-\b}]=0$ if  $\a-\b \in \cI$ or $\a-\b=0$.
One of the two conditions is always satisfied for any pair of roots $\a,\b$ of $\msu(3)$
(notice that the indices were not renamed in rewriting \eq{V-phi-rewrite}).
Therefore this cubic term vanishes identically for the zero modes, which therefore completely decouple 
from the rotating background, as long as interactions with the ``heavy'' excitation modes
can be neglected.

Now we want to see if some of these would-be zero modes $\phi_\a$ 
can acquire a non-vanishing time-independent VEV, denoted as
``non-trivial Higgs'' solutions. We should accordingly obtain the equations of motion
in the presence of such $\phi_\a\neq 0$, separating
the time-dependent $X_\a$ from the time-independent $\phi_\a$. 
Consider first  the variation of the cubic terms:
\begin{align} 
\tr X_\a \d\Box_\phi \phi^\a 
 &= \tr  X_\a \Big([\d\phi_\b,[\phi^\b,\phi^\a]]  + [\phi_\b,[\d \phi^\b,\phi^\a]]\Big)  \nn\\ 
 &= \tr\d\phi_\b \Big(- 2[X_\a,[\phi^\b,\phi^\a]] -  [\phi^\b,[\phi^\a,X_\a]]   \Big) 
 \label{cubic-variation}
\end{align}
using the Jacobi identity. Using again the gauge-fixing condition, it follows
\begin{align} 
\d_\phi V_{\rm int}[\phi]
 &= \frac{m^4}{g^4} \tr\d\phi^\a \Big( 2[[\phi_\a,\phi^\b],X_\b] 
 + \Box_\phi X_\a  + \Box_\phi \phi_\a \Big)  .
 \label{int-variation-simple}
\end{align}
Thus the full equation of motion for our background with non-vanishing Higgs 
 can written  compactly as
\begin{align}
 \big(\frac{\Box_4}{m^2} + \Box_X  +  \Box_\phi + 2\Di_{ad}^\phi  \big) X_\a
 + \big(\frac{\Box_4}{m^2} +  \Box_X + \Box_\phi + 2\Di_{ad}^X \big) \phi_\a
 = 0
 \label{eom-Higgs}
\end{align}
where $\Di_{ad}^X\equiv \Di_{ad}$, and we introduce
\begin{align}
(\Di_{ad}^\phi X)_\a &= \sum_\b [[\phi_\a,\phi^{\b}],X_{\b}] 
=  ((\slashed{D}_{\rm mix}^{\phi}  + \slashed{D}_{\rm diag}^{\phi}) X)_\a \nn\\
(\slashed{D}_{\rm mix}^{\phi} X)_\a &:= \sum_{\b\neq \a} [[\phi_\a,\phi^{\b}],X_{\b}]  \nn\\
(\slashed{D}_{\rm diag}^{\phi} X)_\a &= 2[[\phi_\a,\phi_{-\a}],X_{\a}]  \qquad\mbox{(no sum)}
 \label{higgs-decouple-1}
\end{align}
in analogy to section \ref{sec:fluctuations}.
To maintain the decoupling between the rotating background $X_\a$ and the 
low-energy  Higgs $\phi_\a$, we should therefore require that
\begin{align}
\slashed{D}_{\rm mix}^{\phi} X_\a = 0
 \label{D-vanishing-phi}
\end{align}
in addition to \eq{D6-vanishing}.
Then the above equations of motion take the symmetric form 
\begin{align}
 \big(\frac{\Box_4}{m^2} + \Box_X  +  \Box_\phi +  2\slashed{D}_{\rm diag}^{\phi}  \big) X_\a &= 0  \nn\\
  \big(\frac{\Box_4}{m^2}  +  \Box_\phi +  \Box_X + 2 \slashed{D}_{\rm diag}^{X}  \big) \phi_\a &= 0
   \label{eom-Higgs-2}
\end{align}
For illustration, we  elaborate one non-trivial solution based on the 
lowest zero modes of type \eq{zeromodes-examples}, given by 
\begin{align}
 \phi_\a = \varphi_\a\, T_{-\a} .
 \label{higgs-nonlin-ansatz}
\end{align}
with real-valued $\varphi_\a = \varphi_{-\a}$.
Then $\slashed{D}_{\rm mix}^{\phi}X_\a = 0$ holds for precisely the same reason as 
 $\slashed{D}_{\rm mix}^X\phi_\a = 0$, as discussed in section \ref{sec:zero-modes}.
By literally repeating the computation leading to $M^2_\a$ in \eq{Higgs-mass}, we obtain
\begin{align}
 (\Box_\phi +2 \slashed{D}_{\rm diag}^{\phi}) X_{\pm\a_i} &= M^{X}_i X_{\pm\a_i}, 
\end{align}
 with 
 \begin{align}
 M^{X}_{3} &= 2\big(\varphi_1^2\a_1 + \varphi_2^2 \a_2 - \varphi_3^2 \a_3,\a_3\big)
  = 2(\varphi_1^2 + \varphi_2^2  - 2\varphi_3^2), \nn\\ 
 M^{X}_{2} &= 2\big(\varphi_1^2\a_1 + \varphi_2^2 \a_2 - \varphi_3^2 \a_3,-\a_2\big)
  = 2(\varphi_1^2 -2\varphi_2^2  +\varphi_3^2), \nn\\
 M^{X}_{1} &= 2\big(\varphi_1^2\a_1 + \varphi_2^2 \a_2 - \varphi_3^2 \a_3,-\a_1\big)  
  = 2(-2\varphi_1^2 +\varphi_2^2  +\varphi_3^2)
\label{X-mass}
\end{align}
the equations of motion become 
\begin{align}
 \big(\frac{\Box_4}{m^2} + \Box_X  +  M_{\a}^{X} \big) X_i^\pm &= 0 ,  \nn\\
 \big(\frac{\Box_4}{m^2} + \Box_\phi + M^\phi_\a \big) \phi_i^\pm  &= 0
\end{align}
where $M^\phi_i = M^\phi_{\pm\a_i}$ is the eigenvalue of $(\Box_X + 2\Di_{ad}^X)$ 
given by \eq{M-phi-results}.
Using also \eq{Box-X-explicit} and its analog 
\begin{align}
  \Box_\phi \phi_1^\pm &= 2(2 \varphi_1^2 +\varphi_2^2 + \varphi_3^2) \phi_1^\pm, \nn\\
  \Box_\phi \phi_2^\pm &= 2(\varphi_1^2 + 2 \varphi_2^2 + \varphi_3^2) \phi_2^\pm, \nn\\
  \Box_\phi \phi_3^\pm &= 2(\varphi_1^2 + \varphi_2^2 + 2\varphi_3^2) \phi_3^\pm .
  \label{Box-phi-explicit}
\end{align}
we obtain the eom\footnote{The equations in $\varphi_i$ are actually cubic, and we only display 
the reduced form assuming that all $\varphi_i \neq 0$.}
\begin{align}
 \begin{pmatrix}
  2&1&1&-2&1&1\\
  1&2&1&1&-2&1\\
  1&1&2&1&1&-2\\
  -2&1&1&2&1&1\\
  1&-2&1&1&2&1\\
  1&1&-2&1&1&2
 \end{pmatrix}
\begin{pmatrix}
  r_1^2 \\  r_2^2 \\  r_3^2 \\  \varphi_1^2 \\ \varphi_2^2 \\ \varphi_3^2 
\end{pmatrix} 
 = -\frac 1{2m^2}\begin{pmatrix}
                \omega_1^2  \\  \omega_1^2  \\  \omega_1^2  \\  0 \\ 0  \\  0
                 \end{pmatrix}
\end{align}
Assuming that $\varphi_i^2>0$ for all $i$, the last three equations lead to an inconsistency,
which reflects the sum rule $M_1^\phi+M_2^\phi+M_3^\phi = 0$ \eq{sum-rule}.
Therefore either one or two of the $\varphi_i$ must vanish.
It turns out that there is no solution with two non-vanishing $\varphi_i$, but
there is a solution with $\varphi_3 \neq 0$ and $\varphi_1=0=\varphi_2$,
given by
\begin{align}
 \begin{pmatrix}
  r_1^2\\ r_2^2\\ r_3^2 \\ \varphi_3^2
 \end{pmatrix}
 = \frac{1}{8m^2}\left(
\begin{array}{c}
 -2 \omega_1^2+2 \omega_2^2-\omega_3^2 \\
  2 \omega_1^2-2 \omega_2^2-\omega_3^2 \\
 -\omega_1^2-\omega_2^2+\omega_3^2 \\
 -\omega_1^2-\omega_2^2+2 \omega_3^2
\end{array}
\right) .
\label{nontriv-scalar-solution}
\end{align}
The $r_i^2$ are indeed positive if all $\omega_{i,\mu}$ are approximately equal and timelike,
and
\begin{align}
 \varphi_3^2  &=  \frac 1{8m^2}(2\omega_3^2-\omega_1^2-\omega_2^2) \nn\\
 &= -\frac 14 (r_1^2+r_2^2-2r_3^2) = -\frac 14 M_3^\phi \ > 0
 \label{higgs-VEV}
\end{align}
must also be positive, which is consistent with $M_3^\phi < 0$ \eq{M-phi-results}. 
This holds for suitable frequencies $\omega_i$, which we assume here.
Hence  the scale of the Higgs $\varphi$ is set by 
the {\em difference} of the rotation 
frequencies 
\begin{align}
 m^2\varphi^2 \sim \frac 14\Delta\omega_i^2 \ .
 \label{higgs-VEV-scale}
\end{align}
This is a classical quantity imprinted in the background, which is naturally small.
In principle, this offers an interesting mechanism to introduce a stable hierarchy into the model.
Unfortunately, it turns out that 
the above solutions \eq{nontriv-scalar-solution} have non-vanishing gauge current $\cJ_\mu \neq 0$,
hence a gauge field is induced which will modify the 
background.
Moreover, there are other possible Higgs modes which acquire a negative mass, 
e.g. some of the modes \eq{zeromodes-examples}
\begin{align}
\phi^{\pm (n)}_i &=  Y^{n\a_3}_{\mp n\a_i} \ \sim  \  (T_i^\mp)^n .
\label{zeromodes-explicit-simple}
\end{align}
which will play a role below. 
Therefore a more general ansatz is required, and finding a 
consistent Higgs configuration is a challenging task.
Nevertheless such a mechanism is clearly at work, and it might provide an interesting 
way to introduce scales and a hierarchy into the simple model.

\subsection{Higgs background and Yukawa couplings}
\label{sec:Yukawas}

Now  consider the interaction of the fermionic zero modes with some of the above Higgs zero modes;
we shall also denote them as ``would-be zero modes'' henceforth, since they may acquire 
a mass in the presence of a Higgs. 
The Yukawa couplings  between fermions and Higgs modes arise from  
\begin{align}
 m \Tr \obar\Psi \gamma_5\slashed{D}_{(6)}\Psi
 &= m\Tr \obar\Psi \gamma_5 \Delta^\a[\phi_\a,\Psi] \nn\\
 &= 2m\sum_{i=1}^3 \bar\Psi\gamma_5  \Big(\Delta^-_i [\phi_i^+,.] + \Delta_i^+ [\phi_i^-,.]\Big)\Psi
\end{align}
where $\Psi$ is subject to the Majorana-Weyl condition 
$\Psi^C = \Psi = \Gamma\Psi$.
We focus on the contribution from the fermionic zero modes 
\begin{align}
 \Psi_{\L'} &=  |s_1s_2s_3\rangle\,  Y^\L_{\L'}
\end{align}
as in \eq{fermion-zeromodes}. 
Here $Y^\L_{\L'}$ is an extremal weight state in $V_\L$.
Now assume that due to a negative mass term,
some of the following Higgs modes on $\cC_N[\mu]$ 
are switched on, such as
\begin{align}
  \phi_{i}^{\pm(n)} = \varphi_i^{(n)} \,(T_i^{\mp})^n   
 \label{Higgs-n-explicit}
\end{align}
 as discussed above.
 For suitable $n$ (which may depend on $i,\L$),
these $\phi_{i}^{\pm(n)}$ relate extremal weights 
$\L' \in \cW\L$  of $V_\L$ in adjacent Weyl chambers, 
\begin{align}
 [(T_i^{\mp})^n,Y^\L_{\L'}] \ &\sim  \ w_i \cdot Y^\L_{\L'}  = Y^\L_{w_i\L'} 
\end{align}
where $w_i$ is the Weyl reflection along $\a_i$.
Moreover, $\Delta_i^\pm$ implements the same Weyl reflection of the spinors as defined in \eq{Weyl-spinors},
e.g.
\begin{align}
 \Delta^-_i |\uparrow\uparrow\uparrow\rangle = \tilde\omega_i \cdot |\uparrow\uparrow\uparrow\rangle, \qquad 
 i=1,2,3  \ . 
 \label{Clifford-weyl-relation}
\end{align}
It is then easy to see  that
\begin{align}
 (\Delta^-_i [\phi_i^+,.] + \Delta_i^+ [\phi_i^-,.])\Psi_{\L'} \ \sim \  \varphi_i^{(n)} \tilde\omega_i \cdot \Psi_{\L'}
\end{align}
for fermionic zero modes $\Psi_{\L'}$ and appropriate $n$ (and only for those).
As discussed in section \ref{sec:fermions}, $\tilde\omega_i\cdot\Psi_{\L'}$ is the zero mode $\Psi_{w_i\L'}$
if $w_i\L'$ is adjacent to $\L'$, but not otherwise;
for example if $\L = \a_3$, then 
$\tilde\omega_3\cdot\Psi_{\L} \neq \Psi_{w_3\L} = \tilde\omega_2 \cdot\tilde\omega_3\cdot\tilde\omega_1\cdot\Psi_\L$,
hence $\tilde\omega_3\cdot\Psi_{\L}$ is not a zero mode.
Therefore we obtain Yukawa couplings 
\begin{align}
m \,\tr\bar\Psi\gamma_5 \Di_{(6)}\Psi &\ \sim \ 
2 m \sum_{i,n,\L''= w_i\L'} \tr\bar\Psi_{\L''}\gamma_5\ \varphi_i^{(n)} \tilde\w_i \cdot \Psi_{\L'} 
 \label{Yukawa-zeromodes}
\end{align}
relating adjacent fermionic zero modes with the appropriate $\varphi_i^{(n)}$.
There are also terms coupling the fermionic zero modes to various
heavy fermionic modes, which presumably can be neglected at low energies.

To summarize, extremal fermionic zero modes in adjacent Weyl chambers related by  a Weyl reflection
acquire a mass in the presence of a suitable
 Higgs mode $\langle\varphi_i^{(n)}\rangle \neq 0$,
which links their weights. 
Taking into account the results of
previous section, this mass is expected to be of order 
\begin{align}
 m_\Psi^2 \sim \Delta\omega_i^2 .
 \label{fermion-mass-Higgs}
\end{align}
This will give mass to the off-diagonal standard-model-like fermions in section \ref{sec:standard-model}.
Similar mass terms may arise more generally for  fermionic zero modes
in adjacent Weyl chambers linked by other Higgs fields,
provided the corresponding cubic matrix elements $\tr \bar Y^{\L'}_{w'\L'} [\phi,Y^\L_{w\L}]$
are non-vanishing; this boils down to a  representation-theoretic question.
Thus on a given large $\cC[\mu]$ brane, 
there are many  possible Yukawas coupling the fermionic zero modes via
the various Higgs,
which suggests that they tend to be heavy. The detailed structure is clearly complicated.

\subsection{Fermionic currents and coupling to gauge bosons}
\label{sec:currents}

The interaction of the fermionic zero modes with the gauge bosons 
is given by
\begin{align}
 S_A(\Psi) = \Tr \bar\Psi \gamma^\mu[A_\mu,\Psi]
 =  \Tr J^{\mu \L}_{M}\, A_\mu^{\L,M}
\end{align}
where
\begin{align}
 J^{\mu \L}_{M}(x) &= \bar\Psi(x) \gamma^\mu[Y^\L_{M},\Psi(x)]
\end{align}
and we decompose the gauge field into the $|M,\L\rangle$ modes in $End(\cH_\mu)$,
\begin{align}
 A_\mu(x) = \sum A_\mu^{\L,M}(x) Y^\L_{M}
\end{align}
We are primarily interested in the currents which arise within the low-energy sector,
due to fermionic (would-be) zero modes 
$\Psi_{\L'}$ with extremal weights $\L'$ in $End(\cH)$  as discussed above. 
Since $J^{\mu \L}_{M}$ does not involve any internal Clifford generators, both 
fermions must have the same internal spins, hence their
extremal weights $\L,\L'$ must be in the same Weyl chamber 
(see figure \ref{fig:spin-states-roots}); for example, both must be highest weights.
Thus $J^{\mu \L}_{M} \neq 0$ only if either $M=0$, or if $M$  
relates different extremal weights $\L'$ in the same Weyl chamber.
For the fermions on a given $\cC[\mu]$ brane, this may hold for 
various currents $J^{\mu \L}_{M}$ with $M\neq 0$, if the corresponding cubic matrix elements 
$\tr \bar Y^{\L'}_{w\L'} [Y_M,Y^\L_{w\L}]$ are non-vanishing.
For large branes $\cC[\mu]$, this leads to a complicated structure of
mutually coupled gauge and scalar fields.

However, we will be mainly interested in fermions linking a $\cC[\mu]$ brane with a point
brane, as discussed below. Then there is only one extremal weight $\mu$ in each Weyl chamber,
so that only currents $J^{\mu \L}_{M=0}$ with $M=0$ can arise.
They couple only to gauge bosons  $A_\mu \sim Y^\L_0$ with weight $M=0$. 
The mass of these gauge bosons 
is given by $m_A^2 = 2 m^2 r^2 (\L,\L +2\rho)$ using \eq{gauge-boson-mass},
assuming $r_i \approx r$. 
The most interesting such modes are the  $\chi_L$ and $\chi_R$ modes
identified in section \ref{sec:gauge-bosons}, which  couple to left resp. right-handed fermionic zero modes
$C_L$ resp. $C_R$ of the extremal fermions $\Psi_{\L}$.
This is a crucial ingredient of a chiral gauge theory.

\subsection{Higgs -- gauge bosons coupling}
\label{sec:higgs-gauge}

The interaction of the scalar fields with the gauge bosons 
arises from their kinetic term, via  
\begin{align}
 D_\mu \Phi_i^+ &= m \mu (X_i^+ + \phi_i^+)
  = m(i\omega_{i\mu} X_i^+ + D_\mu \phi_i^+  - i [A_\mu,X_i^+])  \ . 
\end{align}
Keeping only the quadratic terms in the fields, this leads to
\begin{align}
 -\int d^4 x \frac 1{g^2} \tr D_\mu \Phi_i^+ D^\mu \Phi_i^-   
  = -\int d^4 x \frac{m^2}{g^2}\tr\big( \omega_i^2 X_i^+ X_i^- 
    + D_\mu \phi_i^+ D^\mu \phi_i^-  
   + \frac 12 A_\mu \Box_X A^\mu  \big)
 \label{Higgs-gaugebosons-action}
\end{align}
 after some straightforward algebra,
using the decoupling property $[X_i^+,\phi_i^-] = 0 = [X_i^-,\phi_i^+]$ (cf. \eq{decoupl-sum})
as well as $\tr X_i^+  \phi_i^- = 0 = \tr X_i^-\phi_i^+$ 
(since $X_i^+  \phi_i^-$ has non-zero weight).
The first term is the kinetic energy of the rotating background. 
The last term leads to the mass $m_A^2$ \eq{gauge-boson-mass}
of the gauge fields in terms of the eigenvalues of $m^2\Box_X$, as anticipated.
The canonically rescaled scalar fields are
hence given by 
\begin{align}
 \tilde\phi_i^\pm = \frac mg \phi_i^\pm .
\end{align}
As usual, the cubic and quartic terms  will lead to interactions of Higgs 
fluctuations and the gauge fields, and the rotation drops out. 
As discussed in section \ref{sec:rotating-solutions}, there is no linear term $\tr A^\mu \cJ_{\mu}$  since 
we require  $\cJ_\mu =0$, which holds in particular for
$\omega_1=\omega_2 = -\omega_3$ and $r_1 = r_2=r_3$.
In the presence of a non-trivial Higgs VEV $\phi_\a$, the
higher-order terms in $\phi_\a$  must of course be included, 
however this is beyond the scope of this paper.

\section{Stacks of branes}

Now consider fermions in a background which consists of a sum of different branes of the above type,
such as
\begin{align}
 X^a= \begin{pmatrix}
       X^a_{(1)} & 0 \\
       0 &  X^a_{(2)} 
      \end{pmatrix},  \qquad 
\Psi = \begin{pmatrix}
        \Psi_{11} & \Psi_{12} \\
        \Psi_{21} & \Psi_{22} 
       \end{pmatrix} \ .
      \label{point-branes}
\end{align}  
Although a priori the different blocks might have independent parameters, 
stability suggests that the rotation vectors $\omega_i$ are the same for all branes, 
and for simplicity we assume that also the $r_i$ are the same.
Then the above $X^a$ can  be written in the familiar form 
\begin{align}
 X_i^\pm = r_i e^{\pm i \omega_i x}\, T_i^\pm
 \label{X-T-reducible}
\end{align}
where $T^a$ acts on the reducible representation $\cH_{\L_1} \oplus \cH_{\L_2}$.
Some blocks may be trivial, with $\cH_0 = \C$ and
$T^a = 0$, corresponding to point branes located at the origin.
The fermions on the  diagonal blocks behave as discussed above.
On a point brane, there are only trivial fermionic zero modes $\L=0$
with degenerate spin states. The off-diagonal fermions can be written as
$\Psi_{12} = \sum|s_i\rangle |\mu_1,\L_1\rangle\langle\mu_2,\L_2|$, and 
live in  $\cH_{12}= \cH_1\otimes \cH_2^*$.
Similarly $\Psi_{21}$ lives in $\cH_{21}=\cH_2\otimes \cH_1^*$, but this is redundant 
due to the M-W constraint $\Psi_{21} = \Psi_{12}^C$.
Therefore all our results from section \ref{sec:fermions} on fermionic and bosonic zero modes 
carry over directly to the off-diagonal blocks:
The (would-be) zero modes can be collected into 
 $\cN=4$ supermultiplets of trivial $\L=0$ modes,
as well as chiral supermultiplets labeled by
the extremal weights $\L$ in $End(V_{\mu})$. 
On a stack of squashed $\cC_N[\mu_i]$ branes, this leads to a quiver gauge theory,
with gauge group $U(n_i)$ on each vertex $\mu_i$ and 
arrows labeled by $\L$ corresponding to the chiral supermultiplets in $Hom(V_{\mu_i},V_{\mu_j})$
with extremal weights $\L$. These fields transform in the bi-fundamental $(n_i)\otimes (\bar n_j)$
of $U(n_i) \times U(n_j)$.
Charge conjugation relates the modes $\L$ and $\Omega\L$  with reversed arrow, i.e.
$Hom(V_{\mu_i},V_{\mu_j})$ with $Hom(V_{\mu_j},V_{\mu_i})$, so that these fields
are not independent.

The case of a squashed $\cC_N[\mu]$ brane and a point brane is particularly interesting:

\subsection{Point branes attached to squashed brane}
\label{sec:point-attached}

Consider a background consisting of a $\cC_N[\mu]$ brane and a point brane at the origin:
\begin{align}
 X_i^\pm= \begin{pmatrix}
       X_{i(1)}^\pm & 0 \\
       0 & 0
      \end{pmatrix} \ .
      \label{point-branes}
\end{align} 
The first block is given by $End(\cH_1)$ with $\cH_1 \cong V_\mu$ 
 a highest weight module  with highest weight $\mu$.
The off-diagonal  modes live in $\cH_{12} =Hom(\C,V_\mu) \cong V_\mu$
and $\cH_{21} =Hom(V_\mu,\C) \cong V_\mu^*$, with extremal weights 
$\L \in\cW \mu$ resp. $\L \in -\cW \mu$.
This leads to 6 independent off-diagonal fermionic and bosonic zero modes.

Let us discuss the fermionic zero modes $\Psi^{12}_\mu$ in more detail.
Using the above results, they have the form
\begin{align}
 \Psi^{12}_\mu =|\uparrow\uparrow\uparrow\rangle |\mu\rangle_1 \, \psi^{12}_\mu,
\end{align}
along with 5 additional zero modes $\Psi^{12}_{w\cdot\mu}$  obtained by acting 
on $\Psi^{12}_\mu$ with suitable reflections.
Their chirality is determined by the sign $|w|$ of the Weyl group element $w$.
This leads to 3 left-handed fermions denoted by $C_L$, and 3 right-handed fermions 
called $C_R$, as in figure \ref{fig:spin-states-roots}.

Let us estimate the energy scale of the heavy off-diagonal $\Psi^{12}$. Since they all live in 
the irreducible representation $V_\mu$, the mass terms
arising from $m\slashed{D}$ should lead to 
\begin{align}
 m_{\Psi^{12}_{\rm heavy}}^2 \approx \frac 18 |\mu|^2\omega^2
 \label{heavy-offdiag}
\end{align}
which is very large for large branes $\cC[\mu]$.
Then the scale of the off-diagonal heavy fermions is 
much larger than the mass of the relevant gauge bosons in $(0,3)$ resp. $(3,0)$
discussed in section \ref{sec:currents} of order $|\omega|$. 
This is important in the approach to the standard model discussed below.

Now assume that some Higgs modes $\phi_{i}^{\pm(n)} = \varphi_i^{(n)} \,(T_i^{\mp})^n$
on the brane $\cC_N[\mu]$ are switched on.
Using the results of section \ref{sec:Yukawas}, 
this leads to Yukawa couplings \eq{Yukawa-zeromodes} 
relating adjacent fermionic zero modes $\Psi^{12}_\mu$.
Thus the off-diagonal fermionic zero modes  acquire a mass 
in the presence of a suitable
 Higgs mode $\langle\varphi_i^{(n)}\rangle \neq 0$
which links  adjacent modes. No other  fields on $\cC_N[\mu]$ can 
induce such a mass term, and
the couplings between the fermionic zero modes and various
heavy fermionic modes can presumably be neglected at low energies.
Hence the off-diagonal sector is relatively simple.

\subsubsection{Off-diagonal Higgs $S_{12}$}
\label{sec:off-diag-Higgs}

The off-diagonal scalar modes are parametrized as
\begin{align}
 X_i^\pm= \bar X_i^\pm \ + \ \phi_i^\pm 
   &= \begin{pmatrix}
       X_{i(1)}^\pm & 0 \\
       0 & 0
      \end{pmatrix}
   + \begin{pmatrix}
       0 & S_{i(12)}^\pm \\
       S_{i(21)}^\pm & 0
      \end{pmatrix} 
 \label{point-branes-S}
\end{align} 
with $\phi_{i}^\pm = (\phi_{i}^\mp)^\dagger$.
Since $\cH_{(12)}=V_\mu$ and $\cH_{(21)}=V_\mu^*$, we can apply the general results 
in section \ref{sec:zero-modes}:
There are 6 complex scalar (would-be) zero modes
denoted as\footnote{The notation $S$ is used to emphasize the analogy with the $S$ Higgs 
in \cite{Steinacker:2014fja}.}
\begin{align}
  S_{i(12)}^\pm &=  h_i^\pm |\mu_i^\pm\rangle
\end{align} 
with extremal weights $\mu_i^\pm \in \cW \mu$ in the Weyl chamber opposite to $\pm\a_i$.
The other off-diagonal modes are then determined by the hermiticity constraint,
\begin{align}
  S_{i(21)}^\pm &= (S_{i(12)}^\mp)^\dagger = (h_i^\mp)^* \langle\mu_i^\mp| \ .
\end{align}
This has the extremal weight $-\mu_i^\mp$ in the Weyl chamber opposite to $\pm\a_i$ (just like $\mu_i^\pm$),
corresponding to the 6  (would-be) zero modes  
in $\cH_{(21)}=V_\mu^* = V_{\bar\mu}$.
Therefore the off-diagonal low-energy modes correspond to the extremal weights in 
complex representations
$V_\mu \oplus V_{\bar\mu}$, subject to the above hermiticity constraint.

If the $\omega_i$ are different, some of these off-diagonal modes  acquire again a negative mass.
To find solutions with $\langle S_{i}^\pm\rangle \neq 0$,
we can use the results in section \ref{sec:interact-Higgs} for the matrix $\phi_i^\pm$.
The extremal weight properties again imply
$\slashed{D}_{\rm mix}^{\phi} X = 0 = \slashed{D}_{\rm mix}^{X} \phi$, so that 
the eom \eq{eom-Higgs-2}  applies.
In analogy to the solution in section \ref{sec:interact-Higgs},
we assume  that $\phi_{1,2}^\pm = 0$ and
$\phi_3^\pm \neq 0$. However, 
the hermiticity conditions now relates {\em different} off-diagonal modes, and
it is possible to have  $S_{\a(12)}\neq 0$ while $S_{-\a(12)} = 0$;
this will be important in section \ref{sec:standard-model}.
Thus we make the ansatz
\begin{align}
 \phi_3^+ = \varphi\,  \begin{pmatrix}
       0 & |\mu_3^+\rangle \\
       0& 0
      \end{pmatrix}, 
 \qquad \phi_3^- = \varphi^*\,  \begin{pmatrix}
      0 & 0 \\
       \langle\mu_3^+| & 0
      \end{pmatrix}
      \label{S-nonlin-ansatz}
\end{align}
and all other $\phi_i^\pm = 0$.
Then
\begin{align}
 [\phi_3^+,\phi_3^-] 
 = |\varphi|^2 \begin{pmatrix}
                |\mu_3^+\rangle\langle\mu_3^+|  & \\ 
                0 & - 1
               \end{pmatrix} = -  [\phi_3^-,\phi_3^+] 
\end{align}
using $\langle\mu_3^+| \mu_3^+\rangle=1$,
and all other $[\phi_\a,\phi_\b] =0$. 
Therefore the equation of motion \eq{eom-Higgs-2} are 
\begin{align}
 \big(\frac{\Box_4}{m^2} + \Box_X  +  \Box_\phi \big) X_i^+ &= 0 , \quad  i=1,2 \nn\\
 \big(\frac{\Box_4}{m^2} + \Box_X  +  \Box_\phi + 2 \slashed{D}_{\rm diag}^{\phi} \big) X_3^\pm &= 0 ,  \nn\\
 \big(\frac{\Box_4}{m^2} + \Box_\phi +  \Box_X  + 2 \slashed{D}_{\rm diag}^{X} \big) \phi_3^\pm  &= 0 
 \label{off-diag-Higgs}
\end{align}
where
\begin{align}
\slashed{D}_{\rm diag}^{\phi}X_3^\pm  &= \pm 2|\varphi|^2[P_+,X_3^\pm ], \qquad\quad
 P_+ = \begin{pmatrix}
        |\mu_3^+\rangle\langle\mu_3^+| & 0\\ 0 & 0
       \end{pmatrix} ,  \nn\\
   \Box_\phi \phi_3^\pm &= 4|\varphi|^2\phi_3^\pm,     
\end{align}
We also know $(\Box_X  +2 \slashed{D}_{\rm diag}^{X}) \phi_3^\pm = M_3^\phi \phi_3^\pm$ 
from section \ref{sec:zero-modes}, with 
\begin{align}
 M_3^\phi = 2\big(r_1^2\a_1 + r_2^2 \a_2 - r_3^2 \a_3,\mu\big) .
\end{align}
Therefore the last equation in \eq{off-diag-Higgs} would be easy to handle.
On the other hand,  the first equations in  \eq{off-diag-Higgs}  together with
\begin{align}
\Box_\phi X_i^\pm 
  &= 2|\varphi|^2 (P_+ X_i^\pm  + X_i^\pm  P_+)  
\end{align}
require a modification of the matrix elements of $X_i^\pm$. 
The ladder structure of $X_i^\pm$ and hence the crucial decoupling properties
should be preserved, since $P_+$ is diagonal in weight space.
Finding a solution should therefore
not be too hard, but we postpone this to future work.
An additional complication arises in the presence of non-vanishing Higgses $\phi_i^{(0)}$ on  
$\cC_N[\mu]$. 

The important point is that it is possible to have Higgs configurations with 
$S_{3(12)}^+\neq 0$ but $S_{3(12)}^- = 0$,
involving  $|\mu_3^+\rangle$ but not $|\mu_3^-\rangle$. Since  the
 $|\mu_3^+\rangle$  states
are localized on one specific sheet $C_R$ (or $C_L$) of $\cC_N[\mu]$  with specific orientation,
such a configuration  breaks the symmetry between the opposite chiralities, leading
to a chiral low-energy gauge theory with fermions of different chirality behaving differently.
We will briefly discuss in in section \ref{sec:standard-model} how one might obtain in this way
an effective low-energy behavior resembling  the standard model.

\subsubsection{Yukawas for off-diagonal fermions and Higgs}
\label{sec:off-diag-higgs-yuk}

Finally we want to understand the impact of an off-diagonal Higgs $S_{(12)}^a$ as in \eq{point-branes-S}
to the off-diagonal fermions $\Psi^{12}_\mu$.
We compute
\begin{align}
 \big[S_{i(21)}^+,\Psi^{12}_\mu\big] 
 &= (h_i^-)^* \big[\langle\mu_i^-|,|\uparrow\uparrow\uparrow\rangle |\mu\rangle_1\big] \nn\\
  &= (h_i^-)^* \Big(\langle\mu_i^-| \mu\rangle \, \one_2 - | \mu\rangle_1  \langle\mu_i^-|_1 \Big) |\uparrow\uparrow\uparrow\rangle \psi^{12}_\mu \nn\\
  \big[S_{i(12)}^+,\Psi^{21}_\mu\big] &= h_i^+
   \Big(\langle\mu|\mu_i^+\rangle \, \one_2 - | \mu_i^+\rangle_1  \langle\mu|_1 \Big)|\uparrow\uparrow\uparrow\rangle  \psi^{21}_\mu
 \label{S-psi-comm}
\end{align}
and similarly for $\phi_i^-$.
Consider first the Yukawa couplings with the fermionic zero modes $\Psi_{22}$ on 
the point brane, which arise from the terms $\sim \one_2$ above.
Assuming that $\mu$ is in the dominant Weyl chamber,
only  $i=3$ contributes in the Yukawa coupling
\begin{align}
 m\tr \obar\Psi_{22}\g_5 \Big(\Delta_i^- [S_{i(21)}^+,\Psi^{12}_\mu] 
  + \Delta_i^+ [S_{i(12)}^-,\Psi^{21}_\mu\big] \, \Big)
\end{align}
leading to 
\begin{align}
m(h_3^-)^* \langle\mu_3^-| \mu\rangle
 \,\obar\Psi_{22}\g_5 \Delta_3^-  \,|\uparrow\uparrow\uparrow\rangle \psi^{12}_\mu  .
\end{align}
This is non-vanishing, since $\Psi_{22}$ is a trivial $\L=0$ mode.
More generally, the chiral zero mode $\Psi^{12}_\mu$ with $\mu$ 
equal to the weight $\mu_i^\pm$ of some non-vanishing $S_{i(12)}^\pm$ is combined
with the corresponding polarization of $\Psi_{22}$ into a massive Dirac fermion,
which are thereby raised into a massive sector.
If the Higgs configuration is such that $S_{i(12)}^+\neq 0$ but $S_{i(12)}^- =0$, 
then a chiral low-energy gauge theory emerges at low energies comprising only the 
remaining un-paired chiral\footnote{This is not in contradiction with the index theorem, since 
the other chirality is in the conjugate sector, which is however redundant due to the MW condition.} 
modes in $\Psi^{12}_\mu$.
In the context of the standard-model-like configuration discussed below,
this may provide a mechanism to remove $\nu_R$ from the low-energy spectrum.

The second terms in \eq{S-psi-comm} lead to Yukawa couplings of $\Psi^{12}_{\mu'}$
with the fermionic zero modes $\Psi_{11}$ on $\cC[\mu]$.
Assume  that $\mu'=\mu$ is the highest weight in $V_\mu$, we obtain the Yukawas
\begin{align}
 - m (h_i^-)^* \tr \big(\obar\Psi_{11} | \mu\rangle_1  \langle\mu_i^-|_1  \big) \,
 \Delta_i^- \g_5|\uparrow\uparrow\uparrow\rangle \psi^{12}_\mu
\end{align}
Now the $i=3$ contribution can be ignored since then $\mu_3^- = \mu$ and 
$| \mu\rangle_1  \langle\mu_i^-|_1$ has weight zero, but there is no $\Psi_{11}$
zero mode with vanishing weight. For $i=1,2$ the weight
$\mu_i^-$ is adjacent to $\mu$ so that the polarizations match, but the
overlap  $\tr\big(Y^\L_{\L'}|\mu\rangle \langle\mu_i^-|\big)$ with the extremal mode 
in $Y^L_{\L'}\in End(\cH_\mu)$ with $\L'=\mu-\mu_i^-$ may be small. 
Thus there may be some suppressed Yukawa couplings between the 
$\Psi_{12}$ and $\Psi_{11}$ zero modes, which however requires more work.

\section{Towards the standard model}
\label{sec:standard-model}

As an example of the above configurations, consider 
two congruent 6-dimensional branes $\cC_N[\mu]$ denoted by $\cD_u$ resp. $\cD_d$, and
one extra point brane denoted by $\cD_l$.
Let us denote by $\mu_i^R$ 
the 3 extremal weights  related to $\mu$ by a Weyl rotation,
and by  $\mu_i^L$ the remaining 3 extremal weights; in other words, 
$C_L = \{\mu_i^L\}$ and $C_R = \{\mu_i^R\}$ for $i=1,2,3$, cf. figure \ref{fig:spin-states-roots}.
As discussed before, off-diagonal chiral fermionic zero modes $\Psi_{\L}^{(12)}$ arise, connecting  $\cD_l$
with the points $C_L$ and $C_R$ in $\cD_u$ resp. $\cD_d$. 
We denote these off-diagonal fermions as follows:
\begin{align}
 l_L^i = \begin{pmatrix}
  \nu_L^i\\ e_L^i
 \end{pmatrix} \equiv \begin{pmatrix}
                       \Psi^{(ul)}_{\mu_i^L}\\[1ex]
                       \Psi^{(dl)}_{\mu_i^L}
                      \end{pmatrix},
\qquad\quad  l_R^i = \begin{pmatrix}
   \nu_R^i\\ e_R^i
 \end{pmatrix} \equiv  \begin{pmatrix}
                        \Psi^{(ul)}_{\mu_i^R} \\[1ex]
                        \Psi^{(dl)}_{\mu_i^R}
                      \end{pmatrix} .
\end{align}
These play the role of the leptons, with the appropriate chirality. 
Here $i=1,2,3$ corresponds to the generations. 
This background admits an unbroken $U(2)\times U(1)_l$ gauge symmetry,
where $SU(2)$ mixes $\cD_u$ and $\cD_d$ so that
$l_L$ and $l_R$ are doublets. The $U(1)_l$ arises from $\cD_l$.

Now assume that there is an off-diagonal Higgs  $S_\a^{(ul)} \neq 0$ connecting $\cD_l$
with $\cD_u$, attached to one (or several) of the extremal weight states $\mu_i^R$ of $C_R$.
As discussed in section \ref{sec:off-diag-Higgs}, we may and will assume that the corresponding $S_{-\a}^{(ul)}=0$.
This means that $S_\a^{(ul)}$ transforms like  $\nu_R^i$.
Then the above symmetry is broken to $U(1)_{ul}\times U(1)_d$,
where $U(1)_{ul}$ arises on the $\cD_u\cup\cD_l$ branes.
In particular, $\nu_R$ becomes a singlet.

To obtain baryons, we add three extra baryonic point branes $\cD_{b_j}$, $j=1,2,3$.
Now assume that there is no Higgs linking these baryonic branes 
and the leptonic ones\footnote{One possibility to ensure this might be to let the point branes 
rotate in a suitable way;
however it is not clear if this can be consistently implemented.}.
We denote the fermionic zero modes linking $\cD_{b_j}$ with
$\cD_u$ resp. $\cD_d$ according to chirality as 
\begin{align}
 q_L^i = \begin{pmatrix}
  u_L^i\\ d_L^i
 \end{pmatrix} \equiv \begin{pmatrix}
                       \Psi^{(ub)}_{\mu_i^L}\\[1ex]
                       \Psi^{(db)}_{\mu_i^L}
                      \end{pmatrix},
\qquad\quad
 q_R^i = \begin{pmatrix}
   u_R^i\\ d_R^i
 \end{pmatrix} \equiv  \begin{pmatrix}
                        \Psi^{(ub)}_{\mu_i^R} \\[1ex]
                        \Psi^{(db)}_{\mu_i^R}
                      \end{pmatrix} .
\end{align}
The unbroken symmetry can then be organized as $U(1)_Q\times U(1)_b \times U(1)_{tr} \times SU(3)_c$,
where the electric charge generator
\begin{align}
Q = \frac 12\big(\one_{ul} -\one_d -\frac 13 \one_b \big)
\end{align}
gives the correct charge assignment via the adjoint (cf. \cite{Steinacker:2014lma}),
and $U(1)_{tr}$ is the overall trace-$U(1)$.
Here $\one_{ul} = \one_u+\one_l$, and $Q$ is traceless.
The $U(1)_{tr}$ decouples in $\cN=4$ SYM, but it acquires an interesting role related to gravity 
in the IKKT matrix model \cite{Steinacker:2010rh}.

Now consider the lowest massive gauge bosons. As explained in sections 
\ref{sec:currents}, only those gauge bosons with weight $M=0$ couple to the low-energy 
currents arising from the off-diagonal fermions.
In the absence of $S^{ul}$,
the lowest among those are the $M=0$ modes in $(1,1) \otimes \mmu_2$ 
where $\mmu_2$ denotes the $\mmu(2)$ algebra acting on the coincident branes $\cD_u,\cD_d$.
Let us however first discuss
the next-to-lowest modes, given by the weight zero generators in 
$((3,0) + (0,3))\otimes \mmu_2$ discussed in section \ref{sec:gauge-bosons}.
In particular, the mode $\Theta$ \eq{theta-def} measures the chirality of the extremal weight states, 
i.e. $\Theta = \pm c$ on $C_L$ and $C_R$, respectively\footnote{On the point branes, $\Theta$ vanishes.}. 
We can therefore write 
$\Theta = \chi_L - \chi_R$ for the zero modes
 where $\chi_{L,R}$ are characteristic functions on $C_{L,R}$, or equivalently
$2\chi_L = \Theta +c\one$ and $2\chi_R = -\Theta +c\one$.

Now assume that the off-diagonal Higgs  $S^{(ul)}$ is switched on, with a large mass scale. Then  
$\Theta^u$ acquires a mass, while $\chi_L^u$ commutes with $S^{(ul)}$ and therefore survives at low energies.
Together with the second brane $\cD_d$ the remaining low-energy generators are $\chi_L\otimes \msu_2$
which generates $SU(2)_L$, 
and another $U(1)$ generated by $\chi_R^d$. However since the latter is anomalous upon restricting to the 
off-diagonal zero modes\footnote{Assuming that $\nu_R$ disappears from the low-energy spectrum as 
discussed in section \ref{sec:off-diag-higgs-yuk}.
Any anomaly which may arise upon restricting 
to a subset of fermionic zero modes must cancel upon including all modes, since the underlying 
model is anomaly free.}, 
we assume (as in  similar constructions in the context  of intersecting branes \cite{Antoniadis:2000ena})
that $\chi_R^d$ is replaced by the combination 
\begin{align}
 Y &= \one_{ul}  - \frac 13 \one_b - (\chi_L^u + \chi_R^d)\ 
\end{align}
corresponding to weak hypercharge, which is traceless and anomaly free in the off-diagonal low-energy sector.
For the zero modes, this amounts to the Gell-Mann-Nishjima formula
\begin{align}
 2Q-Y = \chi_L^u-\chi_L^d = 2T^3_L \ .
\end{align}
It is  plausible that the mass scale of $Y$ is set by
$\chi_R^d$, and should be comparable to the mass of the $W$ bosons
\begin{align}
 m_Y^2 \approx m_W^2  \sim 18 m^2 r^2  \approx \frac{9}{4} \omega^2 .
\end{align}
Now we come back to the $M=0$ gauge boson modes in $(1,1)$ on $\cC[\mu]$, given by
the  Cartan generators $H_{3,8}^u$. These don't fit into the standard model, because 
they are sensitive to the generation.
Those on $\cD_u$ acquire a mass due to $S_\a^{(ul)}$.
There are candidates for Higgs modes on $\cC[\mu]$ 
(notably the $\L=(3,0)$ and $\L = (0,3)$ modes which relate the weights within $C_L$  resp. $C_R$)
which give mass to all $H_{3,8}$ modes but not to the $\chi_{L,R}$
modes. If those are switched on, then the $H_{3,8}$ modes would become massive,
leaving the above $SU(2)_L \times U(1)_Y$ as lowest gauge bosons. However
this is simply an assumption at this point.
In addition, there is a Kaluza-Klein tower of gauge bosons 
on $\cC[\mu]$ with roughly equidistant masses  above $m_W$. 
They are expected to acquire additional mass terms form 
various Higgses, which hopefully raise them sufficiently far above the $W$ scale.

We have thus identified all the off-diagonal (would-be) fermionic zero modes with standard model
fermions, plus $\nu_R$. They are expected to acquire a mass
of order  \eq{fermion-mass-Higgs}
\begin{align}
 m_\psi^2\sim\Delta\omega_i^2
\end{align}
due to the Higgs fields $\phi_\a^{(0)}$ on $\cD_{u,d}$.
Since in the (illustrative) solution \eq{higgs-VEV} only one 
polarization $\phi_\a^{(0)}$ is switched on, it seems likely that 
one generation is much heavier that the others; this is also suggested by the sum rule \eq{sum-rule}. 
On the other hand, additional fermionic (would-be) zero modes arise on 
the branes $\cD_{u,d}$ and $\cD_{l,b}$.
Some  fermions on $\cD_l$ will pair via $S^{(ul)}$ with $\nu_R$ to form a massive Dirac fermion, 
according to section \eq{sec:off-diag-higgs-yuk}.
The fermions on $\cD_b$ are superpartners of the gluons in a $\cN=4$
supermultiplet. Finally the fermions on $\cD_{u,d}$ 
are mixed by many  possible Yukawa couplings via the various Higgs modes, as discussed in section \ref{sec:Yukawas}.
This suggests that they tend to be heavier than the off-diagonal fermions,
which have a rather clean structure of Yukawas. 
The details are clearly complicated and require much more work.

We also recall 
that for branes $\cC[\mu]$ with large $\mu$, 
the heavy off-diagonal fermions \eq{heavy-offdiag} are much heavier than the $W$ mass. 
This is an important improvement over the  configuration in 
\cite{Steinacker:2014fja}.
While the ``internal'' sector of zero modes on the $\cD_{u,d}$ is much simpler and cleaner
on minimal branes $\cC[(1,1)]$,  
the heavy off-diagonal fermions would be  roughly at the same scale as the $W$ mass.
Hence the case of large $\mu$ seems more attractive from the particle physics point of view.

It is important to note that up to now, we discussed only the ``upper-diagonal'' $\Psi^{12}$-type modes between 
the branes. However this gives indeed the full story, since the conjugate $\Psi^{21}$ modes
are fixed by the MW condition and do not represent independent physical modes.
We emphasize again that it is the assumption that $S^{(ul)}$ is connected to $C_R$ but not $C_L$
which is responsible for obtaining the above chiral standard-like 
behavior, starting from  $\cN=4$ SYM.

We content ourselves with these qualitative observations here, leaving a more detailed 
analysis to future work.
No claim on the viability of such a scenario can be made.
However, given the special status of $\cN=4$ SYM (and the IKKT matrix model),
it is certainly very intriguing that one can arrive quite naturally in the vicinity of 
the standard model (in the broken phase), reproducing all its odd quantum numbers and even
``predicting'' the number of generations.

\section{Energy and current}

Let us compute the energy density of the background, which  in our conventions is 
given by
\[
 E = T_{00} = \frac 1{4g^2}\tr \left( 2  D_0 \Phi^a D_0 \Phi_a 
  + D_\mu \Phi^a D^\mu \Phi_a 
  + \Phi_a \Box_\Phi \Phi_b \right).
\]
Using $r_i = r$ and $\omega_1=\omega_2=-\omega_3 = \omega$  \eq{allequal-constraint} and 
 $\Box_\Phi \Phi_a = -\omega^2\Phi_a$,
we obtain 
\begin{align}
 E = \frac 1{2g^2}\tr D_0 \Phi^a D_0 \Phi_a
  &= \frac 1{2g^2}\omega_0^2\, m^2 r^2\, \tr(\sum_{\a \in\cI} T^\a T_\a)
 \label{Energy-rotating}
\end{align}
using \eq{ri2-omega}. 
Here $\tr(\sum_{\a\in\cI}T^\a T_\a)$ can easily be evaluated in a given representation,
cf. \cite{Steinacker:2014lma}.
This is clearly a  large energy density, set by the UV scale of the model. 
Similarly, the  $R$-current  
corresponding to the  $U(1)_i$ rotation generated by $\t_i$ with frequency $\omega_i$ is
\begin{align}
 J_{\mu}^i &= \frac 1{g^2} \tr\big(\del_\mu \Phi^a (\t_i)_{ab} \Phi^b\big) 
  = \frac 1{g^2} m^2 r_i^2 \omega_{i\mu} \tr(T^{\a_i} T_{\a_i} + T^{-\a_i} T_{-\a_i} )
  \label{J-i-explicit}
\end{align}
This is the same for all $i$ due to \eq{vanishing-J-cond-2},
so that the energy per combined $R$-charge is
\begin{align}
 \frac{E}{|J_0^1|+|J_0^2|+|J_0^3| } = \frac 12 \omega_0 \ . 
\end{align}
%
These expressions for $E$ and $J_\mu$ apply quite generally for 
various rotating brane solutions. For example, there are rotating fuzzy sphere solutions
given by $X_i^\pm = r e^{\pm i\omega x} J_i^{(l)}$ where $J_i^{(l)}$ are the spin $l$  
generators of $\msu(2)$.
All these solutions have the same  energy per $R$ charge ratio for given $\omega$
(the formulae apply even for 
a rotating plane wave solution $X_i^\pm \sim m r e^{\pm i\omega x}$ with light-like 
current $J_\mu \sim\omega_\mu$).
Therefore any of these solutions costs the same energy per given $R$-charge density, and  there is no
clear ``ground state'' with lowest energy per $R$ charge, for fixed frequency $\omega$.

\section{Discussion and conclusion}

We have discussed a new class of generalized vacua of $\cN=4$ SYM
which are not Poincare-invariant but rotating, and introduce a scale into the theory.
Remarkably, they lead to a 
sector of low-energy excitations with  Poincare-invariant kinematics,
which is  oblivious to the background rotation up to interactions with 
the remaining sector of ''deformed`` excitations.  
The backgrounds have a  geometrical interpretation as spinning fuzzy squashed
coadjoint orbits of $SU(3)$  \cite{Steinacker:2014lma} embedded in 6 extra dimension. 
The Poincare-invariant  sector
includes fermionic zero modes with 
distinct chiralities, analogous scalar zero modes, and a tower of gauge bosons.
They are governed by a chiral gauge theory, in the sense that
the chiral fermionic zero modes 
couple to a KK tower of massive gauge fields with charges depending on their chirality. 
In addition to these zero modes, there are  generalized Kaluza-Klein
towers of modes with deformed 
dispersion relations. Most of these are ''heavy``, but we also find 
a set of tachyonic modes or resonances. We have to assume that 
they get stabilized in some way, e.g. by quantum effects.

 Moreover, we observe that 
if the 3 rotation frequencies $\omega_i^2$ of the background are distinct, 
then some of the scalar (would-be) zero modes  acquire a non-trivial VEV, with scale set by the 
{\em difference} of the frequencies. Although a fully consistent such solution remains to be found,
this may provide a  mechanism for introducing a hierarchy of 
scales into the theory.

On stacks of such branes, gauge theories of quiver type  arise, 
retaining part of the underlying supersymmetry structure.
While the detailed content of these low-energy theories is rather complicated
and needs further elaboration,
it can be studied effectively using group-theoretical tools. 
The structure of the Yukawa couplings is investigated, leading to a 
intricate pattern of interactions with the low-energy VEVs
of the various Higgs candidates. 

Finally, we identified simple configurations of stacks of such branes, whose low-energy sector is argued to be 
close to the standard model in the broken phase,  
 reproducing  all the quantum numbers and even  the number of generations. 
The standard model matter fields can be interpreted as strings connecting different branes. 
Additional fields arise, whose detailed structure and physics 
remains to be clarified. Clearly no claim on physical viability can be made at this point,
but it is certainly  intriguing that one can arrive quite naturally in the vicinity of 
the standard model (in the broken phase), starting from pure $\cN=4$ SYM.

The results and constructions of this paper generalize immediately to the IKKT or IIB 
matrix model \cite{Ishibashi:1996xs}, which can be viewed as a non-commutative $\cN=4$ SYM.
While the noncommutative field aspects discussed here should have a smooth commutative limit \cite{Hanada:2014ima},
the trace-$U(1)$ sector acquires an interesting role related to gravity 
\cite{Steinacker:2010rh}.
The present results are therefore also very encouraging towards considering 
matrix models as fundamental theories, 
cf. \cite{Kim:2011cr,Klammer:2009ku,Aoki:2010gv} for interesting recent results. 
Furthermore, given the much-studied relation between $\cN=4$ SYM and string theory on 
$AdS^5 \times S^5$ \cite{Maldacena:1997re}, it is natural to ask about a dual description in terms of 
supergravity or string theory. 
In the present paper, the effective spectral geometry of the extra dimensions was exhibited; its description
could be made more explicit along the lines of \cite{Steinacker:2010rh}.
On the other hand, the relation with supergravity 
should be seen upon taking into account quantum corrections.
Our solutions are also reminiscent of the story of 
giant gravitons \cite{McGreevy:2000cw}, 
which are dual to operators with non-vanishing $R$-charge \cite{Corley:2001zk}.
Therefore the solutions might be interpreted as a condensate of 
some sort of giant gravitons in supergravity, however we leave this to future work.

The present paper proposes an unconventional new approach to obtain interesting
low-energy physics from $\cN=4$ SYM. There are clearly many open issues, and 
in particular it remains to be seen if the stability problem for the 
rotating branes can be resolved in a satisfactory way. However many variations of the 
approach are conceivable, which may bring
$\cN=4$ SYM or its deformations  closer to actual physics than previously thought.

\paragraph{Acknowledgments.}

This work is supported by the Austrian Fonds f\"ur Wissenschaft und Forschung under grant P24713.
I would like to thank in particular J. Zahn for related collaboration and mathematica programs.
This work also benefited from useful discussions with P. Anastasopoulos, H. Aoki, A. Chatzitavrakidis, 
C-S. Chu, H. Kawai, J. Nishimura,  and A. Tsuchyia.

\appendix

\section{Structure  of the potentials and mixing}
\label{app:potential} 

In this appendix, we identify the eigenmodes of the static part of the potential assuming $r_i=r$,
and show that it is strictly positive apart from the zero modes identified before. 
Its eigenstates have definite polarization $\a \in \cI$ and weights $M$ in $\cH_\L$. 
Moreover, we will establish positivity of the operator $\cO_V$ \eq{OV-def} based on results in 
\cite{Steinacker:2014lma}.

It is convenient to collect the fluctuation modes as
\begin{align}
 \phi  \ = \sum_{\a\in\cI} \l^{\a} \otimes \phi_\a \quad \in (8)\otimes \cH_\L
 \label{phi-8-decomp}
\end{align}
for $\a\in\cI$, subject to $\phi^\dagger = \phi$. 
We can restrict ourselves to a given highest weight irrep $V_\L$, with basis $|M\rangle_\L$.
Then the quadratic potential takes a simple form,
separated into static potential
\begin{align}
 V_{\rm stat}[\Phi] = \tr\,\Phi \big(\frac 12\Box_X +  \slashed{D}_{\rm diag}\big)  \Phi 
\end{align}
and rotating potential 
\begin{align}
 V_{\rm rot}[\Phi] = \tr\, \Phi\slashed{D}_{mix}\Phi ,
\end{align}
noting that $\tr\l_\a\l_\b = 2\k_{\a\b}$.
Consider first the static potential. It is not hard to see that in the representation \eq{phi-8-decomp},
we can write $\slashed{D}_{\rm diag}$ as
\begin{align}
 \slashed{D}_{\rm diag} = -([\l_3,.][H_3,.] + [\l_8,.][H_8,.] ) \ .
\end{align}
Clearly, both $\Box_X$ and $\slashed{D}_{\rm diag}$ are diagonal on the product
states $|L,M\rangle \equiv |L\rangle_{(8)}\otimes |M\rangle_\L \in (8) \otimes H_\L$ 
with  weights $L$ resp. $M$. 
Using the assumption  that the radii $r_i = r$ are 
all the same, we can write 
 \begin{align}
  \frac 1{r^2}\Box_X = \Box_T - ([H_3,[H_3,.]] + [H_8,[H_8,.]])
 \end{align}
 where $\Box_T = \sum_{a=1}^8 [T_a,[T_a,.]]$ is the quadratic Casimir on $V_\L$.
Therefore
\begin{align}
 \slashed{D}_{\rm diag}|L,M\rangle
  &= 2(M,L) |L,M\rangle   \nn\\[1ex]
  \Box_X|L,M\rangle 
  &= 2\big((\L,\L +2\rho) - (M,M)\big) |L,M\rangle  
\end{align}
where $\rho=\a_1+\a_2=\a_3$ is the Weyl vector of $\msu(3)$.
 As a consistency check, we recover $\Box_X X^a = 8 r^2 X^a$.
Therefore the $\Phi_{LM} =|L,M\rangle$ states are indeed the eigenstates 
of $\frac 12\Box_X +  \slashed{D}_{\rm diag}$.
On these states, we obtain
\begin{align}
 V_{\rm stat}[\Phi_{LM} ]
 &=  r^2\tr\, \Phi_{LM}^\dagger \Big((\L,\L +2\rho) - (M,M) -(M,2L) \Big) \Phi_{LM}   \nn\\
 &\geq r^2\tr\, \Phi_{LM}^\dagger \Big((\L,\L +2\rho) - (\L,\L) -(M,2L)\Big) \Phi_{LM} \nn\\
 &= 2r^2 \tr\, \Phi_{LM}^\dagger \Big((\L,\rho) -(M,L) \Big) \Phi_{LM} .
\end{align}
This is clearly non-negative, and vanishes only if $(M,L) = (\L,\rho)$ and its 
images under $\cW$.  But these are precisely the extremal zero modes $\phi_\a^{(0)}$ discussed above,
with fixed polarization $\a$.
Thus  $V_{\rm stat}[\Phi]$ is positive for all modes except the 
zero modes $\phi_\a^{(0)}$.

Now consider the mixing operator $\slashed{\obar D}_{mix}$, fox fixed background.
The corresponding bilinear from can be written in the 
following transparent way \cite{Steinacker:2014lma}
\begin{align}
\frac 12\Tr\Phi^\dagger\slashed{\obar D}_{mix} \Phi
 &= \sum_{\a,\b,\c \in\cI} i g_{\a\c\b} \tr(\Phi^{\dagger\a} [X^\b, \Phi^\c]) \nn\\
   &= -\frac 12\sum_{\a,\b,\c \in\cI} \tr(\l_\a [\l_\b,\l_\c]  \Phi^{\dagger\a} [X^\b, \Phi^\c]) \nn\\
   &=  -\frac 12\sum_{\b \in\cI}\tr(\Phi^\dagger [\l_\b, [X^\b, \Phi]]) \nn\\
   &=  -\frac 12\tr(\Phi^\dagger \slashed{\widetilde D}_{mix} \Phi) 
\end{align}
where we introduce\footnote{This form is valid to 
compute the bilinear form, but note that 
$\slashed{D}_{mix} \Phi$ differs from the definition \eq{} by terms with
$\l_8\otimes \Phi_8 + \l_8\otimes \Phi_8$; these drop out in the bilinear form.}
\begin{align}
 \slashed{\widetilde D}_{mix} \Phi &= -\sum_{\b\in\cI} [\l_\b,[X^\b, \Phi]]   \nn\\
 \slashed{\widetilde D}_{ad} \Phi &= -\sum_{b=1}^8 [\l_b,[X_b, \Phi]] 
 = \slashed{\widetilde D}_{mix}\Phi + \slashed{D}_{\rm diag} \Phi
\end{align}
acting on $\Phi\in (8)\otimes \cH_\L$.
This implies that
\begin{align}
 -2\slashed{\widetilde D}_{ad} &= \Box_{T+L} - \Box_T - 6
\end{align}
where
\begin{align}
 L_a = [\l_a,.],  \qquad  \sum_{a=1}^8 L_a L_a = 12 .
\end{align}
We can  simultaneously diagonalize $M$ and $\L$ (but not $\L'$)
acting on $\Phi\in (8)\otimes \cH_\L$.
Then $\slashed{D}_{ad}$ is diagonal on the fluctuation modes corresponding to  
combined irreps with highest weight $\L'$,
\begin{align}
 \Phi \in (8)\otimes V_\L = \oplus V_{\L'} .
 \label{V-L-8-decomp}
\end{align}
with eigenvalues
\begin{align}
 2\slashed{D}_{ad} = - 2(\L',\L'+2\rho) + 2(\L,\L +2\rho) + 12
\end{align}
on $V_{\L'}$.
Thus we can write
\begin{align}
\cO_V =  \Box_X + 2 \slashed{\obar D}_{ad} &= \Box_X -\Box_{T+L} + \Box_T + 12 \nn\\
  &= 2 \Box_T  - \Box_{T+L} - ([H_3,[H_3,.]] + [H_8,[H_8,.]]) + 12 
 \label{OV-explicit}
\end{align}
It is not easy to compute the spectrum of $\cO_V$, since these operators
do not commute. However,
we observe that $\cO_V$ differs from the static potential (B.7) in \cite{Steinacker:2014lma} 
only by the sign of ${\widetilde D}_{mix}$, and the two are in fact unitarily equivalent. Indeed
using $\t {\widetilde D}_{mix} = - {\widetilde D}_{mix} \t$ \eq{tau-D-anticomm}, 
we have
\begin{align}
 \t \cO_V \t =  \Box_X - 2\slashed{\widetilde D}_{mix} + 2\slashed{D}_{\rm diag} \ .
\end{align}
This is precisely the potential considered in \cite{Steinacker:2014lma}, 
which was shown to be positive definite except for the above  zero modes,
and a set of exceptional zero modes with $\L' = m\L_1$ and $\L'=m\L_2$.
This implies that $\cO_V$ is also positive semi-definite.
We can  recover the (regular) zero modes of $\cO_V$ for $\L'=\L+\rho$ and $M=\L$ 
directly from \eq{OV-explicit}, where the operators commute, with  eigenvalue
\begin{align}
 4 (\L,\L +2\rho)  - 2(\L',\L'+2\rho) - 2(M,M)  + 12  = 0 \ .
\end{align}

\end{document}